\documentclass[aoas,preprint]{imsart}

\RequirePackage[OT1]{fontenc}
\RequirePackage{amsthm}
\RequirePackage[numbers]{natbib}
\RequirePackage[colorlinks,citecolor=blue,urlcolor=blue]{hyperref}

\usepackage{latexsym,epsfig,graphicx,amsmath,amssymb,amscd,undertilde,multirow,psfrag,paralist,xcolor,float, subcaption, mwe}

\newcommand{\thetavec}{{\boldsymbol{\theta}}}

\newcommand{\betavec}{{\boldsymbol{\beta}}}

\newcommand{\E}{\mathbf{E}}

\newcommand{\pr}{{\rm Pr}}

\newcommand{\thetavechat}{\widehat{\thetavec}}

\newcommand{\wh}{\widehat}

\newcommand{\Xvec}{\boldsymbol{X}}

\newcommand{\xvec}{\boldsymbol{x}}
\newcommand{\ivec}{{\boldsymbol{i}}}

\newcommand{\bb}{\mbox{\bf b}}

\newcommand{\be}{\mbox{\bf e}}

\newcommand{\bv}{\mbox{\bf v}}

\newcommand{\bx}{\mbox{\bf x}}

\newcommand{\bA}{\mbox{\bf A}}
\newcommand{\bB}{\mbox{\bf B}}

\newcommand{\bD}{\mbox{\bf D}}

\newcommand{\bE}{\mbox{\bf E}}
\newcommand{\bF}{\mbox{\bf F}}

\newcommand{\bI}{\mbox{\bf I}}

\newcommand{\bM}{\mbox{\bf M}}
\newcommand{\bP}{\mbox{\bf P}}
\newcommand{\bQ}{\mbox{\bf Q}}

\newcommand{\bU}{\mbox{\bf U}}
\newcommand{\bV}{\mbox{\bf V}}
\newcommand{\bJ}{\mbox{\bf J}}
\newcommand{\bW}{\mbox{\bf W}}
\newcommand{\bX}{\mbox{\bf X}}

\newcommand{\bone}{\mbox{\bf 1}}
\newcommand{\bzero}{\mbox{\bf 0}}
\newcommand{\bveps}{\mbox{\boldmath $\varepsilon$}}

\newcommand{\bdelta}{\mbox{\boldmath $\delta$}}
\newcommand{\btheta}{\mbox{\boldmath $\theta$}}

\newcommand{\bleta}{\mbox{\boldmath $\eta$}}

\newcommand{\bmu}{\mbox{\boldmath $\mu$}}

\newcommand{\bSig}{\mbox{\boldmath $\Sigma$}}
\newcommand{\bTheta}{\mbox{\boldmath $\Theta$}}
\newcommand{\bGamma}{\mbox{\boldmath $\Gamma$}}
\newcommand{\bLambda}{\mbox{\boldmath $\Lambda$}}
\newcommand{\bkappa}{\mbox{\boldmath $\kappa$}}
\newcommand{\bOmega}{\mbox{\boldmath $\Omega$}}
\newcommand{\bXi}{\mbox{\boldmath $\Xi$}}
\newcommand{\var}{\mbox{var}}
\newcommand{\mE}{{\mathbb E}}
\def\T{{ \mathrm{\scriptscriptstyle T} }}

\def\lbk{\left \{}
\def\rbk{\right \}}

\def\lmk{\left [}
\def\rmk{\right ]}

\def\lsk{\left (}
\def\rsk{\right )}

\arxiv{arXiv:0000.0000}

\startlocaldefs
\numberwithin{equation}{section}
\theoremstyle{plain}

\endlocaldefs

\begin{document}

\begin{frontmatter}
\title{Disentangling  and Assessing Uncertainties in Multiperiod  Corporate Default Risk Predictions}
\runtitle{Uncertainties in Corporate Default Risk}

\begin{aug}
\author{\fnms{Miao} \snm{Yuan}\thanksref{m1}\ead[label=e1]{miaoy89@vt.edu}},
\author{\fnms{Cheng Yong} \snm{Tang}\thanksref{t1,m2}\ead[label=e2]{yongtang@temple.edu}},
\author{\fnms{Yili} \snm{Hong}\thanksref{t2,m1}\ead[label=e3]{yilihong@vt.edu}}
\and
\author{\fnms{Jian} \snm{Yang}\thanksref{t3,m3}\ead[label=e4]{Jian.Yang@ucdenver.edu}}

\thankstext{t1}{Supported in part by National Science Foundation  Grants IIS-1546087 and SES-1533956.}
\thankstext{t2}{Supported in part by the National Science Foundation Grant CMMI-1634867. }
\thankstext{t3}{Supported in part by the National Natural Science Foundation of China  Grant 71571106.}
\runauthor{M. Yuan, C.Y. Tang,  Y. Hong, and J. Yang}

\affiliation{Virginia Tech\thanksmark{m1}, Temple University\thanksmark{m2}, and University of Colorado Denver\thanksmark{m3}}

\address{Department of Statistics\\
Virginia Tech University\\
213 Hutcheson Hall\\
Blacksburg, VA 24060\\
USA\\ 
\printead{e1}\\
\phantom{E-mail:\ }\printead*{e3}}

\address{Department of Statistical Science\\
Temple University\\
1810 North 13 Street\\
Philadelphia,PA 19122\\
USA\\
\printead{e2}}

\address{Business School\\
University of Colorado Denver\\
1475 Lawrence St\\
Denver, CO 80202 \\
USA\\
\printead{e4}
}
\end{aug}

\begin{abstract}
Measuring  the corporate default risk is broadly  important   in economics and  finance. Quantitative methods have been developed to predictively assess future corporate default probabilities.   However,  as a more difficult yet crucial problem,  evaluating the uncertainties associated with the default  predictions remains little explored.
In this paper,   we  attempt to fill this blank by developing  a procedure for quantifying the  level of  associated uncertainties upon carefully disentangling multiple contributing sources.  
Our framework  effectively  incorporates broad information from  historical  default data, corporates' financial records, and macroeconomic conditions  by a) characterizing the default mechanism,  and  b) capturing the future dynamics of  various features contributing to the default mechanism.   Our procedure overcomes the major challenges in this large scale  statistical inference problem and makes it practically feasible by using parsimonious models, innovative methods, and modern computational facilities.
 By  predicting the marketwide total number of defaults and assessing the associated uncertainties, our method can also be applied for evaluating the aggregated market credit risk level.
Upon analyzing a US market data set, we demonstrate that the level of uncertainties associated with default risk assessments is indeed substantial.  More informatively, we also find that the  level  of uncertainties associated with the default risk predictions is correlated  with the level of default risks,  indicating potential  for  new scopes in practical applications including
 improving  the accuracy of default risk assessments.
\end{abstract}

\begin{keyword}[class=MSC]
\kwd[Primary ]{62P25}
\kwd[; secondary ]{62N99}
\end{keyword}

\begin{keyword}
\kwd{Competing Risks}
\kwd{Corporate Default Probability}
\kwd{EM Algorithm}
\kwd{Dynamic Factor Model}
\kwd{High-dimensional Time Series}
\kwd{Prediction Interval}
\end{keyword}

\end{frontmatter}

\section{Introduction}

Measuring the corporate  default risk  has long been crucial in many business decisions. Examples include loan evaluation where a bank analyzes the credit quality of a borrower over various future potential borrowing periods,  internal control considerations where corporate management needs to periodically  and accurately  assess the firm’s present financial condition,  investment screening where investors predict financial health of investments under consideration and screen out undesirable investments,  and determining the credit ratings by rating agencies.

Also, recent introduction and expansion of  credit derivative markets have renewed  interests in this topic. According to the survey by the International Swaps and Derivatives Association (ISDA), the credit default swap (CDS) market, the most popular type of credit derivatives, has exploded over the past decade to about \$30 trillion in 2010, up from \$0.9 trillion in 2001. The default probabilities underlie the pricing of such financial instrument, and CDS reflects the market-based estimate of default probabilities. The Basel II bank regulation has further pushed the topic to the center of the banking regulation. In particular, based on the Basel II accord, banks and bank regulators need to determine the appropriate level of regulatory and economic capital to be held by a bank to be in line with the credit risk represented by its loan portfolio, where borrower default probabilities play an explicit role.

In the  finance literature,  there are two broad categories of approaches  for  corporate  default modeling --  the structural and the reduced form modeling approaches.
The classical structural approach of \cite{Merton1974} assumes that a firm defaults when its assets drop to a sufficiently low level relative to its liabilities. A key implication is that a firm's conditional default probability is completely determined by the only key variable, its distance to default, which is closely  related to the firm's annual asset growth, accounting for its levels of liabilities and volatilities;  see, among others, the review of the structural approaches in \cite{Altman2004}.
\cite{Campelletal_2008_JF} argue that despite  the impressive predictive power of the \cite{Merton1974}'s structural model,  in light of its restrictive functional form, it is  better to use a reduced-form model,  allowing more covariates entering default predictions;  see also \cite{Duffie2001} on the dependence of defaults with other covariates.  The  first generation of the reduced-form models, e.g., those in \cite{Beaver1966}, \cite{Beaver1968} and  \cite{Altman_1968_JF},  primarily rely on the multiple discriminant analysis (MDA),  classifying a firm into one of the possible categories based on  the score and rank computed from its individual characteristics. 
The second generation approaches,   such as \cite{Ohlson1980} and \cite{Zmijewski1984},  propose to use the logistic regression analysis, attempting to assess the conditional probability that a firm would go default in the next period of time.  Both the first and second generations may be considered as static approaches, as they have been using single-period classification or probability models with the bankruptcy data but have ignored their multiple-period feature.  As pointed out by \cite{Shumway2001}, static models would produce biased and inconsistent estimates of bankruptcy probabilities due to their ignoring the dynamics over time, and may introduce an unnecessary selection bias into the estimates.
The current generation of the reduced form models incorporates the dynamics over time by examining the duration of the default events.  \cite{Shumway2001} proposes a hazard function based duration modeling with time-dependent predictors; see also \cite{Chava2004}, \cite{Campelletal_2008_JF}, \cite{Bharath2008}, and \cite{Beaver2012} among others.    In particular, \cite{Chava2004} demonstrate the superior predicting performance of \cite{Shumway2001}'s  model over the first (i.e., \cite{Altman_1968_JF}) and second (i.e., \cite{Zmijewski1984}) generations of models.   The  most recent development of the reduced form modeling approaches on default predictions has an emphasis on the corporate defaults over  multiple periods; see, for example, \cite{DuffieSaitaWang2007} and  \cite{DuanSunWang2012}.

While various quantitative procedures have been developed for predicting corporate default probabilities,   point predictions are serving as the dominating measures in the current state of knowledge. A blank in the literature is that the point predictions are equipped with no assessment of the associated prediction uncertainties.
A main reason behind  
 is that the task is too challenging due to the  huge scale and high complexity of the problem.
Clearly, the historical corporate and macroeconomic time series data are of   high dimensionality and complexity.
Meanwhile, all companies exposed to future default risks would require assessments of  their default risks and  the associated level of uncertainties.
The level of complexity would substantially increase further if one concerns the prediction with multiple future periods.
Fundamentally, there are  multiple contributing sources to the  uncertainties  including  those from  the default mechanism, the future dynamics of the corporates and economic environment, and the model estimation errors; see Section  \ref{s2} for more details  on the source of uncertainties.   

Our investigation intends to develop a procedure  obtaining prediction intervals for quantifying  the level of uncertainties associated with default risk predictions,  taking  the three aforementioned sources of uncertainties into account.
Studies on the prediction intervals with duration modeling have been documented  in the literature in areas such as the reliability; see, for example, \cite{HongMeekerMcCalley2009} and reference therein.  Nevertheless,   existing methods do not apply to the scenario of corporate default prediction due to the unique  challenging practical aspects of the problem;   
 see Sections \ref{s2} and \ref{s3}.

Given the complexity of the models in this scenario,  explicit formulas  generally do not exist for constructing valid prediction intervals.  Thus, our framework resorts to resampling procedures built upon parametric stochastic models.
When the number of companies at default risk is at the order of tens of thousand with history of tens of years,   challenges are arising from 1) complicated structure of covariates
with large number of  unknown parameters, 2) large scale of the data sets, and 3) complicated data structure.  
For example,  the data set in our analysis in Section \ref{sec:app2} for the US market over the period 1990 -- 2009 contains more than 10,000 companies, and the number of monthly observations  
exceeds 1,000,000.   However,  only few companies have observations  during the entire period because many companies  either went default or exited the market due to other reasons.
Among all companies in the data set,   missing data are overwhelming  and the time horizons for those observations are highly heterogeneous among the companies.
Our  Section \ref{s3}  provides  detail on our  framework  developed for uncertainties assessments with prediction intervals for both point predictions and total defaults predictions,     overcoming those challenges by using parsimonious models,  innovative  and computationally efficient methods, and powerful computational facilities.

The proposed framework  in this study will contribute to the literature from several important aspects. First and foremost, compared with the current practice of default probability prediction which typically yields only the point estimate, the introduction of default prediction intervals dramatically improves our understanding and knowledge especially for model diagnosis and statistical inferences. Furthermore, 
a distinguished feature of our measure is  allowing  for not only multiple sources of uncertainties but also the asymmetric nature of default probability prediction intervals so that the lower bound of default probability prediction would not go below zero (which is obviously not sensible).
Appropriately  quantifying the  associated uncertainties is the  key to valid statistical inference on the future default probabilities.  For example, to assess how well their model of default prediction performs,  \cite{Campelletal_2008_JF} compare the fitted point estimate of probability of failure (which is the average of such estimates from each company) with the actual default rate in the market and conclude (p.2916) that their model somewhat overpredicts failures in 1974 to 1975, underpredicts for much of the 1980s, and then overpredicts in the early 1990s. Obviously,  additional scope of the problem may be provided  if the prediction intervals are taken into consideration.
Also, 
the availability of uncertainty quantification for default probability prediction would enable us to further conduct effective forecasting evaluations, e.g., along the line of \cite{DieboldMariano_1995_JBES}, where one would examine whether the apparent improvement of forecast accuracy is statistically significant.  
In our data analysis reported in Section \ref{sec:app2}, facilitated by the prediction intervals,  we are able to show that the out-of-sample aggregated predictions for the total number of defaults work quite well for multiple years.

Moreover, the uncertainties of default probability prediction should be crucial in improving our understanding of default risk pricing on financial markets, and may provide a new venue of exploring distress risk and/or credit risk in asset pricing.  For example,
\cite{Dingetal_2012_JASA}  document the puzzling negative relationship between stock returns and default risk as measured by default probability.   \cite{Gieseckeetal_2011_JFE}  report a puzzling finding on the US corporate bond market that credit spreads are roughly twice as large as default losses and do not respond to realized default rate.
The missing uncertainties of default probability predication could be important.  
As an illustration,  we demonstrate by our data analysis in Section \ref{sec:app2} that the assessments of  uncertainties associated with predicted default probabilities for individual companies  are  indeed highly informative.  First of all, the level of uncertainties can be high, especially for those companies with high predicted default probabilities.   Second, and more interestingly,  we found that by incorporating the width of the prediction interval in a logistic regression for the binary variable defined as a company going  default or not, significant interaction is found between the width of the prediction interval and the point  default probability prediction. This shows that the level of uncertainties associated with the point default probability prediction can be informative practically for solving problems.
Additionally, the uncertainties of default probability prediction should shed light on many important issues in finance where default/credit risk plays a central role.
For example,  \cite{GieseckeKim_2011_MS} explore the systemic risk of the financial sector defined as the conditional probability of failure of a sufficiently large proportion of financial institutions.
We show in our Section \ref{sec:app2} that  our procedure is capable of equipping prediction intervals with aggregated  predicted total number of defaults, a feature that can benefit various studies of systemic risks.

The rest of this paper is organized as follows.  In Section \ref{s2}, we disentangle the sources of uncertainties contributing to the default predictions.
We present  in Section \ref{s3} our framework for predictions and assessing their associated levels of uncertainties  for  future default risks of individual companies and the total number of future defaults on the market.
Section \ref{sec:app2} comprehensively analyzes a large-scale US market data by developing default probability predictions and quantitatively assessing their associated level of uncertainties.
A simulation study evaluating the accuracy of our method is presented in Section \ref{sim}.
Section \ref{sec:con2} concludes this paper and draws the picture for future research. The Supplementary
Material \citep{Yuanetal2018} contains the detail of the EM algorithm in our method. 

\section{Sources of Uncertainties in Default Risk Predictions } \label{s2}

\subsection{Stochastic Time-to-Event}
\label{dis}

We construct our procedure in a setting for multiperiod corporate default risk prediction.   For the default mechanism, we focus on the recent reduced form models for the durations of the defaults and  other competing  risk events.
Event duration modeling can be broadly classified into the
 time-to-event data analyses,  a subject that has been intensively studied  in areas such as reliability in engineering studies, and survival
analysis (see, for example, \cite{meekerescobar1998}, and \cite{KalbfleischPrentice2002}).
The key device for the duration modeling is the event intensity function. In the survival or time-to-event analysis,  intensity function modeling plays a central role;  see the monograph \cite{KalbfleischPrentice2002} and reference therein.   For corporate defaults in finance, predictions with the intensity function model also involve the stochastic nature of the explanatory variables, and we refer to \cite{Duffie2011} as an overview.

Modeling the durations with intensity function  treats the time when a  company defaults  in future 
as a
random variable.
Meanwhile,  to accommodate the fact that a company may exit the market  before going default due to events other than   bankruptcy, e.g., being acquired by another company,   incorporating the competing risks are  required in  modeling the time-to-exit of the companies.
Generally speaking, suppose there are $K$ types of credit  events competing with each other so that only one event that occurs the first is observed.
 Let $T_k$ be the time to the event of type $k$ $(k=1,\dots, K)$.  
We denote the event intensity function by $\lambda_k(t)$ ($\lambda_k(t)\ge 0, t\ge 0$)  for the $k$th type of event.
The event intensity function connects to the survival  function $S_k(t)=P(T_k>t)$  by
$S_k(t)=\exp\left\{-\int_0^t \lambda_k(u) du\right\}$;  see, among others, \cite{KalbfleischPrentice2002} for more detail on  the intensity function.

Additional to observing that the intensity  $\lambda_k(t)$ is a function of time, it is natural to expect that features including the financial healthiness, profitability, growing perspective and etc. are  affecting the future default occurrences.  Meanwhile, the macroeconomic conditions also have impact on  future defaults.
From the predictive perspective, the company specific features and the macroeconomic conditions are also stochastic, a source that also contributes to the uncertainties in the corporate  default predictions.   Therefore, adequately incorporating the dynamic features  is crucial in both predicting the defaults and assessing the associated level of uncertainties.  

We describe here how to incorporate the dynamic features  as explanatory covariate information in the intensity function. Modeling and  estimating  the stochastic covariate will be discussed in Section \ref{sec:cov.process}.
We denote by  a random vector $\xvec_t=(x_{1t},\dots, x_{pt})^{\T}$  indexed by time $t$ containing generic covariates of dimensionality $p$.  
The observed covariate process is then  $\xvec(t_1,t_2)=\left\{\xvec_s: t_1< s\leq t_2\right\}$ representing available covariate information from  $t_1$ to $t_2$.
Subsequently, the intensity function for event type $k$ ($k=1,\dots,K$) for a company at time $t$  with covariate $\xvec_t$ is characterized  by $\lambda_k(t;\xvec_t)$. Such an intensity function 
models the rate (i.e., probability per unit time) that event $k$ will happen instantly after time $t$, conditioning on the covariate value. The total  events  intensity (i.e., something happens) for a company at time $t$ is $\lambda(t;\xvec_t)=\sum_{k=1}^K\lambda_k(t;\xvec_t)$ by  noting that competing risk events are mutually exclusive. We also define the cumulative intensity  function for the event type $k$ as $\Lambda_k[t;\xvec(0, t)]=\int_0^{t}\lambda_{k}(s;\xvec_s)ds$, ($k=1,\dots,K$). Then the aggregated cumulative intensity  function  is $\Lambda[t;\xvec(0, t)]=\sum_{k=1}^K\Lambda_k[t;\xvec(0, t)]$.

\subsection{Parametric Intensity Function and Its Estimation } \label{sec:time.to.event.para.est}

For practical applications,  parametric  intensity functions $\lambda_k(\cdot)$  $(k=1,\dots, K)$ are often assumed for effectively analyzing time-to-event data with meaningful practical interpretations.  In our work, we consider that the intensity function   of event type $k$ at time $t$ has the exponential additive form
$\lambda_k(t;\xvec_{t})=\exp(\beta_{k0}+\beta_{k1}x_{1t}+\cdots+$ $\beta_{kp}x_{pt})$.
Treating $\xvec_t$ as random, the framework is referred  to as doubly stochastic in  \cite{DuffieSaitaWang2007} for corporate default predictions.
The parameter $\betavec=(\beta_{10},\dots,\beta_{1p},\dots, \beta_{K0},\dots, \beta_{Kp})^\T$ is unknown  and needs to be estimated from historical defaults data.  Therefore,  uncertainties associated with the parameter estimation contribute to the uncertainties in the default predictions.

We now describe the maximum likelihood (ML) method for estimating the parameter in the intensity function.
For each company, the time-to-event data are denoted by $\{t_i,\bdelta_{i},\xvec_{i}(0, t_i)\}$  $(i=1,\dots,n)$, where $n$ is the number of companies. Here $t_i$ is the event time for company $i$ if one of the $K$ events happens, and $t_i$ is the last observation time $\tau$ if no event occurred during the data collection period. The event indicator for company $i$ is $\bdelta_i=(\delta_{1i},\dots,\delta_{Ki})^\T$, where $\delta_{ki}=1$ and $\delta_{li}=0, l\neq k$ if  event $k$ happens to company $i$, and $\delta_{li}=0,l=1,\dots,K$, if  no event happens until the last observation time $\tau$.  Last, the observed covariate history from the time origin to $t_i$ for company $i$ is denoted as $\xvec_{i}(0, t_i)=\left\{\xvec_{i,s}: 0< s\leq t_i\right\}$, with $\xvec_{i,s}$ representing the covariates of company $i$ at time $s$.
In this investigation,
we consider $K=2$ types of events hereinafter,  i.e., a company defaults ($k=1$) or exits the market due to other events ($k=2$).

We note that the cumulative distribution function (cdf) of the random variable $T$ for  the time-to-event of a company, given its covariate history $\xvec(0, t)$, is
$F_T(t)=P(T\leq t)=1-e^{-\int_0^t\sum_{k=1}^2\lambda_k(s;\xvec_s)ds}=1-e^{-\Lambda[t;\xvec(0, t)]}$.  The marginal cdf of time to event type $k$, denoted as $T_k$, is
$F_{T_k}(t)=1-e^{-\int_0^t\lambda_k(s;\xvec_s)ds}=1-e^{-\Lambda_k[t;\xvec(0, t)]}$. The probability density function (pdf) of $T_k$ is
$f_{T_k}(t)=\lambda_k(t;\xvec_t)e^{-\Lambda_k[t;\xvec(0, t)]}$.
To differentiate different types of observed events, let $\Delta_k, k=1,2$ be the event indicators. That is, $\Delta_1=1, \Delta_2=0$ if it is a default,   $\Delta_1=0, \Delta_2=1$ if it is an exist  due to other reason, and $\Delta_1=\Delta_2=0$ if no event occurred by the latest observation time  (denoted by $\tau$) in the data set.  Due to two types of competing risks, the observed time-to-event  of a company is therefore $T=\min(T_1,T_2)$.
The  fraction of  failing probability due to the type $k$ event is
\begin{align*}
F_k(t)=&\pr(T\leq t, \Delta_{k}=1)=\pr(T_k\leq
t,T_l>T_k;\text{ for all }l\neq k)\\
=&\int_{0}^{t}f_{T_k}(t_k)\prod_{l\neq
k}\left[1-F_{T_l}(t_k)\right]dt_k=\int_{0}^t\lambda_k(s;\xvec_s)e^{-\Lambda[s;\xvec(0, s)]}ds.
\end{align*}

The joint likelihood of the event times or the last observation times $t_i$'s of  the $n$ given the covariate processes $\xvec_{i,t_i}$ at $t_i$ and covariate history $\xvec_{i}(0, t_i)$ $(i=1,\cdots,n)$ is then given by
\begin{align}
\nonumber
L_T(\betavec|\textrm{DATA})&=\prod_{i=1}^{n}\left(\left(\,\,\prod_{k=1}^2 \left\{\lambda_k(t_i;\xvec_{i,t_i})e^{-\Lambda[t_i;\xvec_{i}(0, t_i)]}\right\}^{\delta_{ki}}\right)\right.
\\
&\times \left.\left\{e^{-\Lambda[t_i;\xvec_{i}(0, t_i)]}\right\}^{\prod_{k=1}^{2}(1-\delta_{ki})}\right), \label{equ:likelihood}
\end{align}
where $\lambda_k(t;\xvec_t)$ is proportional to the probability that a company has an event of type $k$ between time $t$ and $t+dt$, where $dt$ is an infinitesimal amount of time,  $e^{-\Lambda[t;\xvec(0, t)]}$ gives the probability of observing a company survives to time $t$. The parameters $\betavec$ are then estimated by maximizing the joint likelihood of the event times in (\ref{equ:likelihood}).
%
%
In practice, the covariate history $\xvec_{i}(0, t_i)$ for company $i$ is only discretely observable. Therefore, integration of the intensity function of event type $k$ (i.e., $\int_0^{t_i} \lambda_k(s; \xvec_{i,s}) ds$) can be   approximated by  discretization. 

We remark that the uncertainties associated with the parameter estimation can be  conventionally quantified using the standard errors by inverting  the observed Fisher information matrix.  In  the literature, this type of standard errors are typically reported as a measure of level of uncertainties. 
However,
considering  only uncertainties in the parameter estimations is not yet adequate  for assessing the uncertainties associated with the multiperiod corporate default predictions. The reason is that the parameter estimation procedure is a static one conditioning on the covariate process so that it fails to incorporate any future dynamics in the covariate process.

\subsection{Covariate Process and Discrete-time Observations}
\label{cm:review}

The intensity function, on one hand, by its definition is a function of continuous time. On the other hand,  those covariates used for modeling the intensity function can only be observed at a grid of discrete time points.   Thus, the survival function, which relates to the integration of the intensity function,  is typically approximated by taking the intensity function to be piece-wise constant between two adjacent observation times.    Such an approximation in fact relates the intensity function modeling to multiperiod binary response variable regression analysis with the logit or other link functions; see, for example, \cite{Shumway2001}  and  \cite{DuffieSaitaWang2007}.    Moreover,  the one-period ahead  default predictions using the logistic regression type of approaches can also be viewed as a result of piece-wise constant approximation of the intensity function;  see,  among others, the studies of \cite{Hillegeist2004} and \cite{Bharath2008}.   Clearly, the accuracy the approximation is worsen with longer time interval between two observations, and so is the accuracy of the prediction using the logistic regression type approaches.    Therefore, multiperiod corporate default predictions require a more accurate account for the future dynamics; see \cite{DuffieSaitaWang2007} and  \cite{DuanSunWang2012}.

In constructing our procedure,
the dynamics of the covariates process  clearly play an important role in corporate default predictions.  For example, the multiperiod approach of \cite{DuffieSaitaWang2007} relies on a parsimonious high-dimensional vector autoregressive time series model for the covariate process, and numerical approximation for integrations with respect to the future dynamics is needed for assessing the multiperiod future corporate default probabilities.
Our model in this investigation for the high-dimensional vector time series incorporates an autoregressive component for capturing the predictive information  in the conditional mean function. The autoregressive structure is a benchmarking one and has broad applications in time series analysis; see the overviews in the monographs \cite{Tsay_2010}, \cite{Durbin2012} and \cite{Tsay2013}. To exploit further the systematic/structural dynamics among the the covariates, we  further equip the innovations  of the vector time series  with a dynamic factor model that disentangles the contributions from a systematic factor component and an idiosyncratic component.

In  statistics and  financial econometrics, the dynamic factor models are advantageous for its parsimony and  have demonstrated their promising predictive performance in broad areas;  see the monograph \cite{Durbin2012} for an overview.
In the credit risk modeling literature, the dynamic factor models have been demonstrated effective and have many successful applications.   Among them, for example,  a dynamic factor model is applied in conjunction with the intensity function modeling with parametric baseline hazard in \cite{Koopman2008a} for credit rating transitions, incorporating dynamic frailty  accounting for the dependence between defaults.
\cite{Koopman2011} investigate the dynamic frailty-correlations between defaults in different segments and groups of firms, and demonstrate the impact from the latent dynamic factors.
 Equipping the latent factors with attractive practical interpretations such as those effects from the macro, industrial, regional factors and etc.,  the dynamic models is capable of incorporating multiple sources of contributions to assessing the default risks.   \cite{Koopman2012} consider the dynamic factors models incorporating multiple effects for the default counts modeling for the 2008 credit crisis.   More recently, \cite{Schwaab2017} investigate global credit risk concerning multiple  countries, and demonstrate the impacts from the country and industry factors.

An appealing feature of the dynamic factor models  is their convenience in computations. For Gaussian models, the Kalman filter can be conveniently applied, which is the case for our dynamic factor model.  The Kalman filter is also capable of handling missing data  and mixed frequency data;  see  \cite{Braeuning2014} for a recent investigation on forecasting with mixed frequency data.  When additional non-Gaussian observations are incorporated with the dynamic models, the importance sampling based techniques are demonstrated useful for estimations.  As examples, the defaults counts  data are handled together with other covariate in the dynamic factor models of \cite{Koopman2008},  \cite{Koopman2011}, and \cite{Schwaab2017}.


The main objective of our covariate process modeling concerns the dynamic for corporate defaults predictions at the individual firm level, with a combination of the autoregressive and dynamic factor structures.  Our approach integrates the covariate model  with the dynamic intensity function model  outlined in Section \ref{dis},   based on which we conduct the multiperiod corporate default predictions.   Here our attempt targets at the corporate default prediction at the firm level, and it is different from predicting the default counts as in  existing studies, for example,  \cite{Koopman2011}, \cite{Koopman2012}, and  \cite{Schwaab2017}.  Computationally, it is clearly a more demanding objective.
Hence exploiting the impacts and benefits from the latent dynamic factors is clearly desirable in this scenario.
Our approach with the modeling device for  corporate default predictions has a few new methodological features of their own interests.  First, the systematic contributions from the latent factors among the firms are exploited in conjunction with the autoregressive structure in a dynamic intensity function for capturing the dynamics in predicting corporate defaults.
Second, the separated treatments of the covariate process and the default mechanism allow feasibility and convenience for multiperiod predictions for individual firms with the dynamic factor model -- only computationally more efficient Kalman filter is involved when handling high-dimensional time series from thousands of firms.

\subsection{Parametric Stochastic Covariate Process and Its Estimation}\label{sec:cov.process}

We now describe the parametric model for the covariate process considered in our framework.
Specifically,
we denote by $\mathbb X_t$, $t=1,\dots, \tau$ the observed covariate process including both firm-specific covariates for all the companies and the macroeconomic covariates at time $t$.
The firm-specific and macroeconomic covariates, serving as effective reflection of the profitability as well as leverage ratio of assets to debts of a company, and indicators for the economic condition of the nation, are used to model the default risks.
Inspired by  \cite{DuffieSaitaWang2007}, we focus on to  two firm-specific variables -- the distance to default ($D_{i,t}$) and the trailing one-year stock  log-return ($V_{i,t}$) for company $i$ at time $t$.
Here, the distance to default \cite{Merton1974}  is a classical measure in corporate credit risk analysis especially from a structural model point of view.
Roughly speaking, the distance to default is defined as the number of standard deviations of annual asset growth by which the log asset level exceeds the firm's log liabilities. In the classical model of  \cite{Merton1974},  a company's conditional default probability is completely determined by its distance to default.
In our studies,  we use  the distance to default calculated by the method proposed in \cite{DuanSunWang2012}.
For macroeconomic variables, we choose the trailing one-year return on the S\&P 500 index ($S_t$) and the three month Treasury bill rate $(r_t)$.  
%
%
Hence,  we have $\mathbb X_t=(\bD_t^\T,\bV_t^\T,r_t,S_t)^\T$, $t=1,\cdots,\tau$ where $\bD_t=(D_{1,t},\dots, D_{n,t})^\T$, $\bV_t=(V_{1,t},\dots, V_{n,t})^\T$, and $\tau$ is the total number of time points. That is, $\mathbb X_t$ is $m\times1$ vectors where $m=2n+2$ and $n$ is the number of firms.
In the data set for our studies, the observations are available monthly.
To adjust for evident quarterly seasonal effect in the time series,  we take a difference of order 3, and resulting in  a new $m$-dimensional vector time series $\bX_t=\mathbb X_{t+3}-\mathbb X_t$ $(t=1,\dots, \tau')$, where $\tau'=\tau-3$.

Modeling  the dynamics of $\bX_t$ is the most challenging task in default predictions and assessing the associated level of uncertainties, because of the fact that $\bX_t$ is of very high dimensionality.  Take, for example, the US market,  the total number of companies  has exceeded 10,000 since 1990.  Moreover, in an active market, new companies are almost continuously formed while many existing ones exit the market for various reasons, resulting highly un-balanced observations of the time series, i.e., the origin and end times of the components in $\bX_t$ are different with possible missing data for some periods of time.
Furthermore,  the behaviors of components in $\bX_t$ are expected to inter-related with each other in some complicated ways. Thus jointly modeling the tremendously high-dimensional time series becomes daunting, while  further dedicated effort is also necessary for developing methods of parameter estimation and assessing the associated level of uncertainties.

We consider a highly parsimonious  time series model for $\bX_t$ with two key components  specifically for default predictions: a) a mean-reverting autoregressive structure  in the conditional mean of $\bX_t$ given prior observations, and b) a dynamic factor model for the innovations to capture the correlations among the components in $\bX_t$.
We refer to \cite{Tsay_2010} as an introduction for modeling vector valued time series, and \cite{PanYao_2008_Bioka} and \cite{LamYao_2012_AOS} for recent development of factor models for multivariate time series.

Specifically, the conditional mean model is  a modified version of the one considered  in \cite{DuffieSaitaWang2007}:
\begin{align}
\bX_t-\bmu=\bTheta (\bX_{t-1}-\bmu)+\bveps_t, \quad t=2,\dots, \tau'.\label{eq:covmod}
\end{align}
Model (\ref{eq:covmod}) is  a vector auto-regression model mainly to capture the conditional dependence with the mean reverting effects of all the covariates.  The coefficient matrix $\bTheta$ is designed in a parsimonious way following \cite{DuffieSaitaWang2007} as 
\begin{align*}
\bTheta=\begin{pmatrix}
\bkappa_D & \bzero & \bb & 0\\
\bzero& \bkappa_V&\bzero & 0 \\
\bzero &\bzero & \kappa_r & 0 \\
\bzero & \bzero &0 &\kappa_S
\end{pmatrix},
\end{align*}
where $\bkappa_D=\kappa_D \bI_n$, $\bkappa_V=\kappa_V \bI_n$, $\bb=b\bone_n$. $\bI_n$ is an $n\times n$ identity matrix and $\bone_n$ is an $n$-dimensional vector taking value $1$ for all of its elements. Here  we define the mean reverting vector as $\bkappa=(\kappa_D, \kappa_V, \kappa_r, \kappa_s, b)^\T$. The first four elements $\kappa_D, \kappa_V, \kappa_r, \kappa_s$ of $\bkappa$ capture the mean reverting effects of the four selected covariate processes. The current distance to default is modeled jointly by the mean reverting of the previous value, and the effect of departure of Treasury bill rate $r_{t-1}$ from its mean at the previous month through the parameter $b$.

To further capture the  serial and cross-sectional dependence between components in $\bX$,  we propose to apply the following dynamic  factor model (DFM) for the innovation vector $\bveps_t$: 
\begin{align}
\bveps_t&= \bLambda \bF_t + \be_t, \label{eq:facmodel}  \\
\bF_{t}&=\bA \bF_{t-1}+\bleta_t,\quad t=2,\dots,\tau', \label{eq:statem}
\end{align}
where the latent factor $\bF_t$ is a $q\times 1$ vector following an auto-regression process with order 1 (i.e.,VAR(1)).
The principal component analysis  (PCA) is a convenient device for the  factor model structure (\ref{eq:facmodel}); see, for example, \cite{stock2002}.
The dynamic factor structure introduced by   (\ref{eq:statem}) has demonstrated its merits in numerous credit risk modeling; see the discussions in \ref{cm:review}. The state space methods \cite{Durbin2012} are convenient for handling the model setting with (\ref{eq:facmodel}) and  (\ref{eq:statem}).
Here, $\bleta_{t}$'s are assumed to be independently and identically distributed (iid) normal random vectors from $\textrm{N}(\boldsymbol{0}, \bQ)$, for some positive definite matrix $\bQ$. Here $\bLambda$ is a $m\times q$ matrix of factor loadings, and $\bA$ is a $q\times q$ matrix of autoregressive coefficients. The random vectors $\bleta_t$ and $\be_t$ are independent normal random vectors. The covariance matrix of $\be_t$ is assumed to be a diagonal matrix $\bP$.  
Here the factor  model with loading $\bLambda$ is high parsimonious by the fact that the number of common factor $q$  is usually very small, which drastically reduces the number of parameters in the covariance matrix of $\bveps_t$.
As in our data analysis,  the number of factor is chosen as $q=2$ by using the method of  \cite{BaiNg2008}.  Facilitated by the dynamic factor model,  the future dynamics of the covariate process can be effectively incorporated in default predictions and uncertainties assessments.

We collectively denote by $\thetavec_{\bx}=\{\bmu, \bTheta,\bLambda, \bA,\bP,\bQ\}$ all parameters in the covariate proces.  We develop an  expectation-maximization (EM)  algorithm for estimating parameters in dynamic factor model specified by  (\ref{eq:facmodel}) and (\ref{eq:statem}), whose detail is given in the Supplementary Material.   Specifically,  our EM algorithm  efficiently incorporates the hidden factor $\bF_t$ in this tremendously large scale problem with high-dimensional time series and highly un-balanced observations.
In our EM algorithm,    both the E-step and M-step can be conveniently executed for practical implementations.  Most remarkably,    the matrices inversions in our EM algorithm only involves those of size $q\times q$, making it most computationally efficient and feasible for this large scale default prediction problems.

\section{Predictions and Uncertainties Assessments} \label{s3}

\subsection{Procedures for Future Default Predictions}\label{sec:pred}

In our study, predicting the future corporate default probabilities given the available current information is the key objective.
For different levels of interests such as assessing the overall level of credit risks,  one may also need to  predict the total number of defaults  for  the overall market system and certain market sectors.

Let us begin with   the  method for individual corporate default predictions, and then the method for aggregated default predictions.

With the intensity model and the covariate model described in Sections  \ref{sec:time.to.event.para.est} and \ref{sec:cov.process}, the future default probability of company $i$ within $s$ time units in future after the last observation time $\tau$ is,
\begin{align}\nonumber
\rho_i(s;\thetavec)&=\E\left\{\left.\pr[\tau<T_i\leq
\tau+s,\Delta_{1}=1|T>\tau]\right | {\cal F}_{\tau}\right\}\\\label{eqn:rho.i}
&=\E \left\{ \left.\left[\int_{\tau}^{\tau+s}\lambda_1(t;\xvec_{i,t})
\exp\lsk-\left\{\Lambda[t;\xvec_{i}(0,t)]-\Lambda[\tau;\xvec_{i}(0,\tau)]\right\}\rsk dt\right]\right| {\cal F}_{\tau} \right\},
\end{align}
where $\thetavec=(\thetavec_T^\T,\thetavec_{\bx}^\T)^\T$ contains the parameters of the covariate model $\thetavec_{\bx}$ and those of the time-to-event model $\thetavec_T$.  
Here, the expectation is understood as conditioning on the information up to time $\tau$,  denoted by ${\cal F}_\tau$, and $\Delta_1=1$ indicates that it is the default probability of interest so that the corresponding intensity function $\lambda_1(\cdot)$ and the cumulative hazard function are involved in (\ref{eqn:rho.i}).
%
%
We note that $\rho_i(s;\thetavec)$ is a sub-distribution function because $\rho_i(\infty;\thetavec)<1$ due to the existence of other type of competing events.

Because no simple analytical expression for the expectation in \eqref{eqn:rho.i} is available, we use a Monte Carlo simulation approach to evaluate $\rho_i(s;\thetavec)$. The following algorithm is for computing $\hat \rho_i(s;\thetavechat)$ with the estimate $\thetavechat$ from the methods described in  Sections  \ref{sec:time.to.event.para.est} and \ref{sec:cov.process}.
\\[1ex]
\noindent\textbf{Algorithm 1:}
\begin{enumerate}
\item Simulate   $\bX^{\ast}(\tau',\tau'+s)$, the future differenced covariate processes,  for all the firm-specific and macroeconomic covariates from their distribution $\bX(\tau',\tau'+s)|\bX_{\tau'}$  as specified in Section \ref{sec:cov.process}.  
\item By inverting the differencing operator,   the simulated covariate processes  denoted by $\mathbb X^{\ast}(0,\tau+s)$ is re-constructed from the combined differenced data $\bX^{\ast}(0,\tau'+s)=\{\bX(0,\tau'),\bX^{\ast}(\tau',\tau'+s)\}$, where $\bX(0,\tau')$ and $\bX^{\ast}(\tau',\tau'+s)$ are respectively the historical and simulated  future data.
\item For each company $i$,  numerically compute
\begin{align*}
\rho_i^{\ast}(s;\thetavechat)&=\int_{\tau}^{\tau+s}\lambda_1(t;\xvec_{i,t}^{\ast})
\exp\lsk-\left\{\Lambda\left[t;\xvec_{i}^{\ast}(0,t)\right]-\Lambda\left[\tau;\xvec_{i}(0,\tau)\right]\right\}\rsk dt.
\end{align*}

\item Repeat steps 1-3 $M$ times to obtain $\rho_i^{\ast m}(s;\thetavechat), m=1,\cdots,M$ where $M$ is the pre-specified number of the simulation replications.
\item The prediction of $\rho_i(s;\thetavec)$ is obtained by $\hat\rho_i(s;\thetavechat)=M^{-1}\sum_{m=1}^M\rho_i^{\ast m}(s;\thetavechat).$
\end{enumerate}
In the first step of \textbf{Algorithm 1}, to forecast the differenced covariate process, we 
calculate the following for $s=1,2,\cdots$
\[
\tilde \bveps_{\tau'+s}=\widehat \bLambda \widehat \bF_{\tau'+s}+\hat \be_{\tau'+s}, \text{ and } \bX_{\tau'+s}=\hat{\bmu}+\widehat{\bTheta} \bX_{\tau'+s-1}+\tilde \bveps_{\tau'+s},
\]where $\widehat\bF_{\tau'+s}$ is forecasted from the fitted DFM  for $\widehat\bF_t$ and $\hat\be_{\tau'+s}$ is drawn from $\textrm{N}(\boldsymbol{0},\widehat\bP)$.

As for  the point prediction for the aggregated number of defaults in the market coverage of interest,   
%
%
%
let $N_s$ be the cumulative number of defaults at $s$ time units after the last observation time $\tau$ and $RS(t)$ be the set  collecting companies  of interests that are at risk of default at time $t$.  It follows that $N_s=\sum_{i\in RS(\tau)}I_i(s)$ and
$I_i(s)\sim\text{Bernoulli}[\rho_i(s;\thetavec)].$ The point prediction for $N_s$ is $\wh{N}_s=\sum_{i\in
RS(\tau)}\hat \rho_i(s;\wh\thetavec)$.

Though both the point predictions for $\rho_i(s)$ and $N_s$ are informative for measuring future default risks, they do not reflect the uncertain nature of the predictions that we have discussed earlier. In what follows, we describe how to assess the uncertainties associated with the predictions.

\subsection{Assessing Uncertainties at the Aggregate Level}\label{agi}
Prediction intervals (PIs) are used to quantify uncertainty in prediction of future random quantities. Let $N_s$ be the cumulative number of events at a future point. A $100(1-\alpha)\%$ PI for $N_s$ is defined as
$\Pr\left(\utilde{N_s}\leq N_s\leq \widetilde{N_s}\right)=1-\alpha$.
To assess the uncertainties associated with the predicted number of defaults,  a natural choice is to supply a PI denoted by $\left[\utilde{N_s},\,\widetilde{N_s}\,\right]$.
A  naive (plug-in) PI  for this purpose is obtained by solving
\begin{eqnarray}\label{eqn:naive.PI}
F_{N_s}(\utilde{N_s};\wh\thetavec)=\frac{\alpha}{2},\quad \text{ and }
\quad F_{N_s}(\widetilde{N_s};\wh\thetavec)=1-\frac{\alpha}{2}.
\end{eqnarray}
Here $F_{N_s}(n_s;\thetavec), n_s=0,1,\cdots, n'$ is the cdf of $N_s$ where $n'$ is the number of companies in the $RS(\tau)$,  $1-\alpha$ is the desired coverage probability. Note that $N_s$ is a sum of non-identically distributed Bernoulli random variables. An explicit form for $F_{N_s}(n_s;\thetavec)$ is
\begin{align}\nonumber
&F_{N_s}(n_s;\thetavec)= \\\label{eqn:cdf.fnk}
&\frac{1}{n'+1}\sum_{l=0}^{n'}\left\{\frac{1-\exp[-\ivec\omega l(n_s+1)]}{1-\exp(-\ivec\omega l)}\prod_{i\in RS}[1-\rho_i(s;\thetavec)+\rho_i(s;\thetavec)\exp(\ivec\omega l)]\right\},
\end{align}
where $\ivec=\sqrt{-1}$ and $\omega=2\pi/(n'+1)$. The cdf in \eqref{eqn:cdf.fnk} is obtained from a discrete Fourier transform of the characteristic function of $N_s$, which can be viewed as a generalization of the binomial distribution  for a collection of firms with homogeneous default probability. 
We refer to \cite{Hong2013} for more details on the derivation and an efficient implementation for computing $F_{N_s}(n_s;\thetavec)$. Alternatively, one can use some approximation methods such as the ordinary normal approximation or normal approximation with second order correction (e.g., \cite{Volkova1996}).

The PI in (\ref{eqn:naive.PI}) ignores the uncertainties in $\thetavechat$. Thus the coverage probability is generally smaller than the nominal $1-\alpha$ level. These PIs can be calibrated to improve the coverage probability. We will use resampling method by parametric bootstrap to do the calibration.

Using the predictive distribution in \cite{LawlessFredette2005}, a $100(1-\alpha)\%$ PI for $N_s$, denoted by $\left[\utilde{N_s},\,\widetilde{N_s}\,\right]$, is obtained by
\begin{eqnarray}\label{eqn:cali.PI}
\utilde{N_s}=v_{\alpha/2}\quad \text{ and }
\quad \widetilde{N_s}=v_{1-\alpha/2}.
\end{eqnarray}
Here $v_{\alpha}$ is the $\alpha$ lower quantile of random variable $N_s^{\ast}$ specified by distribution $F_{N_s}(\cdot;\wh\thetavec)$ in which $\thetavechat$ is also treated as a random variable. In practice, $v_{\alpha}$ can be computed by simulations. That is, $v_{\alpha}$ is approximated by the $\alpha$ sample quantile of $N_s^{\ast b}, b=1,\cdots, B$. Specifically, we obtain $N_s^{\ast b}$ from $F_{N_s}(n_s;\wh\thetavec^{\ast b})$ given $\wh\thetavec^{\ast b}=(\thetavechat_T^{\ast b\T},\thetavechat_{\bx}^{\ast b\T})^\T$, in which $\thetavechat_T^{\ast b}$ was simulated from  $\textrm{N}(\thetavechat_T,\Sigma_{\thetavechat_T})$ and $\thetavechat_{\bx}^{\ast b}$ was estimated from the simulated covariate processes.

The simulation procedure based on parametric bootstrap is as follows.
\\[1ex]
\noindent\textbf{Algorithm 2:}
\begin{enumerate}
\item Simulate the differenced covariate processes $\bX^{\ast}(1,\tau')$ from the model (\ref{eq:covmod}), (\ref{eq:facmodel}) and
(\ref{eq:statem}) with $\widehat \btheta$.  For each company, the differenced observations at the first month are kept and we do not extrapolate any period with no observations in the original data set. 
\item Re-estimate parameters in the covariate model $\thetavechat_{\bx}^{\ast}$ 
based on the simulated processes through the EM algorithm in  the Appendix. 
\item Take a random sample of $\thetavechat_T^{\ast}$ from its asymptotic distribution $\textrm{N}(\thetavechat_T,\Sigma_{\thetavechat_T})$, where $\thetavechat_T$ and $\Sigma_{\thetavechat_T}$ are estimated from the observed data by the methods in Section \ref{sec:time.to.event.para.est}.
\item With the simulated data $\bX^{\ast}(1,\tau')$ and the new parameter estimates $\thetavechat^{\ast}=(\thetavechat_T^{\ast\T},\thetavechat_{\bx}^{\ast\T})^\T$, \textbf{Algorithm 1} is implemented to predict the default probabilities $\rho_i^{\ast}(s;\thetavechat^{\ast}), i=1,\cdots,n$.
\item Take a random sample $N_s^{\ast}$ from its distribution (\ref{eqn:cdf.fnk}) with parameter values $\thetavechat^{\ast}$.
\item Repeat steps 1 to 5 $B$ times to obtain $N_s^{\ast b}, b=1,\cdots, B$.
\item The $100(1-\alpha)\%$ calibrated PI for $N_s$ is $\left\{N_s^{\ast ([(\alpha/2) B])},\,\, N_s^{\ast ([(1-\alpha/2)B])}\right\}$, where $N_s^{\ast (\cdot)}$ is the ordered version of $N_s^{\ast b}$ and $[\,\cdot\,]$ is the round function.
\end{enumerate}

The fundamental rationale of the above algorithm and the one in the next section is to incorporate all sources of uncertainties as discussed earlier, i.e., those from the stochastic default mechanism, the stochastic covariate process, and the parameter estimation procedures.

\subsection{Assessing Uncertainties for  Corporate Default Probabilities}\label{sec:indi.pred}

By applying the Algorithm 1, one may predict the multiperiod ahead default probabilities for individual corporate for evaluating the future default risk.  Clearly, all sources of uncertainties are contributing in the point default probability estimations.  For example, the final point prediction is the average of a range of possible default probabilities, and the different level of variations among those re-generated default probabilities are reflecting different levels of uncertainties associated with point default probability predictions.
For reflecting the level of uncertainties associated with the default probability predictions, we propose to construct calibrated prediction intervals  based on historical data for companies that are at risk  incorporating all contributing sources of uncertainties.


The procedure for constructing the prediction interval is based on  a large scale parametric bootstrap, similar to the one for the aggregated defaults prediction. Specifically, to incorporate the uncertainties in parameter estimation, in each iteration of resampling, we first simulate the differenced processes  from the fitted covariate model and estimate parameters using the simulated data.
To incorporate uncertainties associated with the parameter estimation of the time-to-event model,  we re-generate model parameters from the estimated joint asymptotic distributions.
Finally,  we simulate multiperiod ahead values of the covariate process given  the last observation in the historical data with the re-estimated parameter.
Then, for each replication of the resampling procedure and for each at risk company,   one multiperiod ahead default probability can be obtained. By repeating the procedure a number of times, we obtain the distribution of the predicted default probabilities and construct the prediction interval correspondingly.
More specifically, we have the following algorithm.
\\[1ex]

%

\noindent\textbf{Algorithm 3:}
\begin{enumerate}
\item Simulate the differenced covariate processes $\bX^{\ast}(1,\tau')$ from the model (\ref{eq:covmod}), (\ref{eq:facmodel}) and
(\ref{eq:statem}) with $\widehat \btheta$.  For each company, the differenced observations at the first month are kept and we do not extrapolate any period with no observations in the original data set.
\item Re-estimate parameters in the covariate model $\thetavechat_{\bx}^{\ast}$ 
based on the simulated processes through the EM algorithm in  the Appendix. 
\item Take a random sample of $\thetavechat_T^{\ast}$ from its asymptotic distribution $\textrm{N}(\thetavechat_T,\Sigma_{\thetavechat_T})$, where $\thetavechat_T$ and $\Sigma_{\thetavechat_T}$ are estimated from the observed data by the methods in Section \ref{sec:time.to.event.para.est}.
\item With the simulated data $\bX^{\ast}(1,\tau')$ and the new ML estimates $\thetavechat^{\ast}=(\thetavechat_T^{\ast\T},\thetavechat_{\bx}^{\ast\T})^\T$, \textbf{Algorithm 1} is implemented to predict the default probabilities $\rho_i^{\ast}(s;\thetavechat^{\ast}), i=1,\cdots,n$.
\item Repeat steps 1 to 4 $B$ times to obtain $\rho_i^{\ast b}(s), b=1,\cdots, B$.
\item The $100(1-\alpha)\%$ PI of default probability for the $i$th company at $s$ time units after the last observation time $\tau$ is $\left\{\rho_i^{\ast ([(\alpha/2) B])}(s),\,\, \rho_i^{\ast ([(1-\alpha/2)B])}(s)\right\}$, where $\rho_i^{\ast(\cdot)}(s)$ is the ordered version of $\rho_i^{\ast b}(s)$ and $[\,\cdot\,]$ is the round function.
\end{enumerate}

\section{US Corporate Default Data Analysis}\label{sec:app2}

\subsection{Data Overview}

We now illustrate an application of our prediction framework on a US Corporate data set containing observations  from January 1990 to November 2009.
The data set contains defaults and other credit events information
of the United States (US) public firms  together with their stock market data from the CRSP (i.e., The Center for Research in Security Prices) database and accounting data from the Compustat database. The entire data set has around 12,000  US companies  and more than 1,000,000 firm-specific monthly observations.
Handling the entire data set is difficult given our limited resource,  and it could take a very long time because we need to perform reasonable number of replicated studies for validation and assessment purposes.
 So we choose a subset of the data with three industrial sectors -- electronic product manufacturers, holding and investment offices, and business services.  
These three industrial sectors contain 3,271 firms of the US market and have experienced a majority number of  defaults.  
Specifically, among  the 3,271 companies on the market from January 1990 up to November 2009, 164 defaulted and 2,049 exited due to other reasons, leaving 1,058 companies at risk at the end of November 2009.
Time-to-event information is available in the data set in terms of the occurrence time of an event and its type which is used to determine if it is a default or an exit due to other reasons.
For the firm specific covariate, we consider the distance to default ($D_t$) and the trailing one year stock return ($V_t$) of each company following our discussion in Section \ref{sec:cov.process}.
To incorporate  macroeconomic conditions, we consider monthly data of the trailing one-year S\&P 500 return ($S_t$) and the three-month Treasury bill rate  ($r_t$)  as covariates as well.

\subsection{Model Estimations}\label{sec:app.para.est}

By applying the proportional hazard model introduced in Section \ref{sec:time.to.event.para.est} using covariate described in Section \ref{sec:cov.process},    the default and other exit intensity functions are modeled by $\lambda_k(t;\xvec_{t})=\exp(\beta_{k0}+\beta_{k1}D_{t}+\beta_{k2}V_t+\beta_{k3} r_{t}+\beta_{k4} S_t$ where $k=1,2$ are respectively corresponding to defaults and other exits. We estimate parameters in the time-to-event model by maximizing the log likelihood function as described in Section \ref{sec:time.to.event.para.est}.
Standard errors for the ML estimates are calculated by inverting the observed information matrix. The point estimates and 95\%  confidence
intervals based on asymptotic normality are given in Table \ref{tab:tab.mle.a73}. 


From Table \ref{tab:tab.mle.a73},  we see negative $\hat \beta_{11}$ but positive $\hat \beta_{21}$,  indicating that lower risk is associated with larger value of distance to default,   but in that case the firm is associated with higher chance of exit the market due to other reasons.  Negative $\hat \beta_{12}$ and $\hat\beta_{22}$ show that higher stock return implies lower risk for both default and other forms of exits which may be due to  that  the trailing one-year stock return is an important indicator for a company's profitability.    As for the effect of macroeconomic variables,  negative $\hat \beta_{13}$ and $\hat \beta_{23}$ indicate that an increase in three-month Treasury bill rate manifests lower risks  for both default and other exits, demonstrating the impacts on credit events  from the overall economic condition of the environment.
As for the effect from the trailing one-year return of the S\&P 500 index upon controlling the level of other covariates,
an increase in its value is  associated with higher  default risk,  but its impact on other exits is not statistically significant.  
The negative impact might due to the correlations between the individual stock returns and the S\&P 500 index; see also the discussion in \cite{DuffieSaitaWang2007}.

\begin{table}
\begin{center}
\caption{ML estimates for parameters and their asymptotic standard errors based data over January 1990 to December 2008.}\label{tab:tab.mle.a73}
\begin{tabular}{crrrrccrrrr}\hline\hline
\multicolumn{5}{c}{Default}&&\multicolumn{5}{c}{Other Exits}\\\cline{1-5}\cline{7-11}
\multirow{2}{*}{Para.}&\multirow{2}{*}{Est.} &\multirow{2}{*}{SE}&\multicolumn{2}{c}{ 95\% CI}&&
\multirow{2}{*}{Para.}&\multirow{2}{*}{Est.} &\multirow{2}{*}{SE}&\multicolumn{2}{c}{ 95\% CI}\\\cline{4-5}\cline{10-11}
&&& Lower & Upper&&&&& Lower & Upper\\\hline
$\beta_{10}$&-6.9126  & 0.2018 & -7.3081 &  -6.5171  &&$\beta_{20}$&  -5.2646 & 0.0666 & -5.3950 & -5.1341 \\
$\beta_{11}$&-0.6803 & 0.0867 & -0.8502  & -0.5105 &&$\beta_{21}$&   0.0504 & 0.0084 & 0.0339  &  0.0669 \\
$\beta_{12}$&-1.1467 & 0.0646 & -1.2734 & -1.0200  &&$\beta_{22}$&  -0.3295 & 0.0401 & -0.4081 & -0.2509\\
$\beta_{13}$&-0.3091 & 0.0542 & -0.4153 & -0.2028  &&$\beta_{23}$&  -0.0450 & 0.0160 & -0.0763 &  -0.0137 \\
$\beta_{14}$& 1.9431 & 0.3974 &  1.1642 &  2.7219  &&$\beta_{24}$&  -0.0839 & 0.1404  & -0.3590 &  0.1913\\    \hline\hline
\end{tabular}
\end{center}
\end{table}


To predict future dynamics of the covariates, we apply the  dynamic factor model specified by (\ref{eq:covmod}), (\ref{eq:facmodel}), and (\ref{eq:statem}) described in Section  \ref{sec:cov.process}.
To estimate the parameters,  
  we apply our EM algorithm discussed in Section  \ref{sec:cov.process} whose detail is given in the Supplementary Material.

The mean reverting parameters in (\ref{eq:covmod}) are estimated as  $\hat\bkappa=(0.63766, 0.63551$, $0.89208,$  $0.63546, -0.00714)^\T$ with estimated standard errors $(0.0031, 0.0049$, \\$0.0343, 0.0333, 0.0176)^\T$ .  The parameter $\bA$  in the vector  autoregressive model (\ref{eq:statem}) for the hidden factors $\bF_t$ is estimated as
$$\widehat\bA=\begin{pmatrix}
 0.3734 & 0.2144\\
 -0.0599 & 0.4803
 \end{pmatrix},$$
whose entry wise estimated standard errors are $0.0955$, $0.1368$, $0.0364$, $0.0598$ for $\hat A_{11}$, $\hat A_{12}$, $\hat A_{21}$, and $\hat A_{22}$ where $\hat A_{ij}$ is the $ij$th component of $\widehat \bA$.
The EM algorithm also returns estimations of the loading matrix $\bLambda$ and covariance matrix $\bP$ of $\be_t$ in (\ref{eq:facmodel}) containing many components that are omitted here.

\subsection{Total Defaults Predictions and Uncertainties}\label{sec:app.popu.pred}

In what follows, when we are conducting  multiple-period predictions,  only data up to the origin of predictions are used when applying our prediction framework.  For each prediction period, the actual observed defaults and covariate processes after the origin of predictions were held out  and only used for validation and assessments afterward.     For example,  observed data from January 1990 to December 2008 were used to predict the default risks  one year ahead in 2009 and so forth.

\begin{figure}
\begin{center}
\begin{tabular}{cc}
\includegraphics[width=0.45\textwidth]{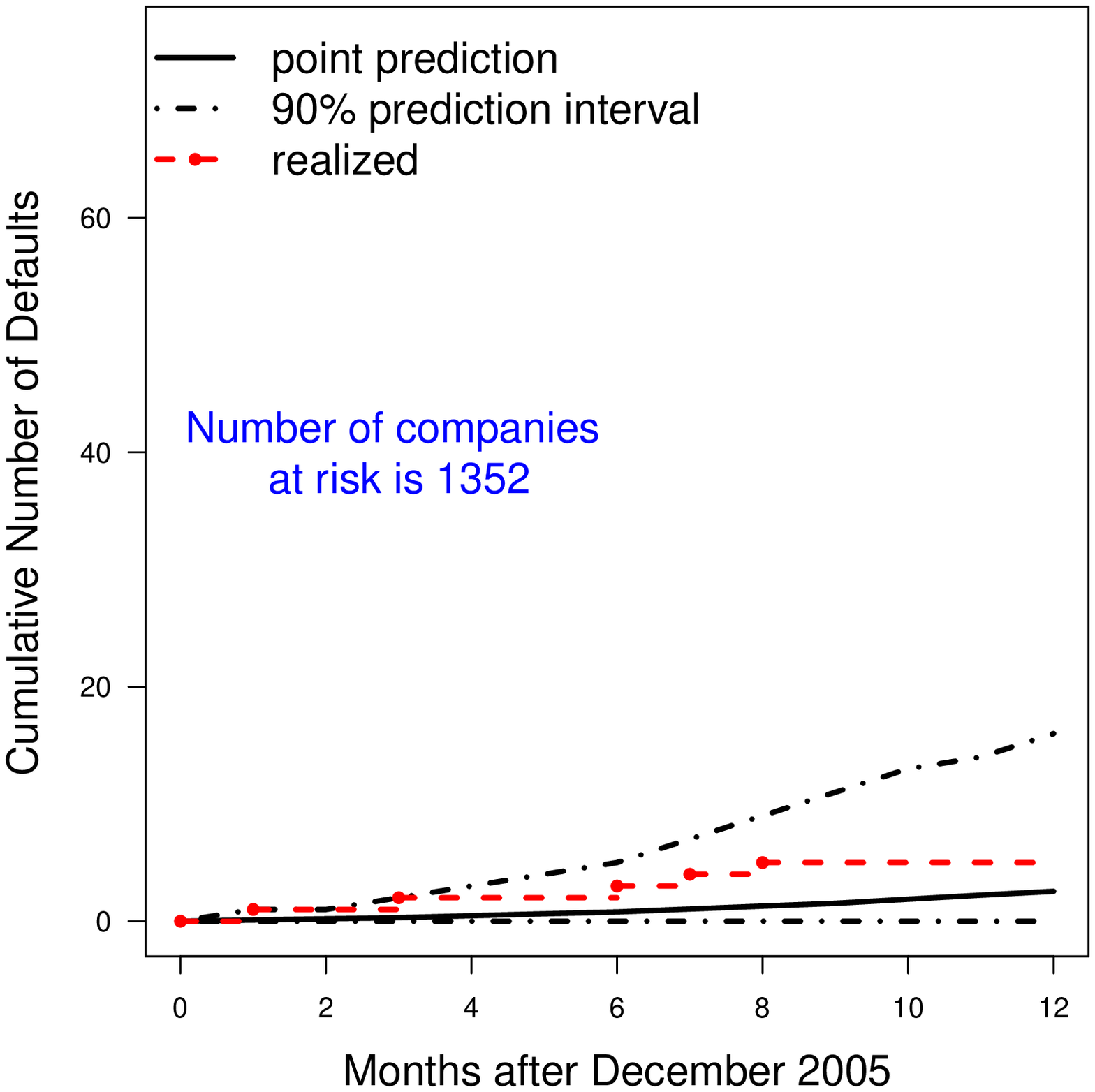}&
\includegraphics[width=0.45\textwidth]{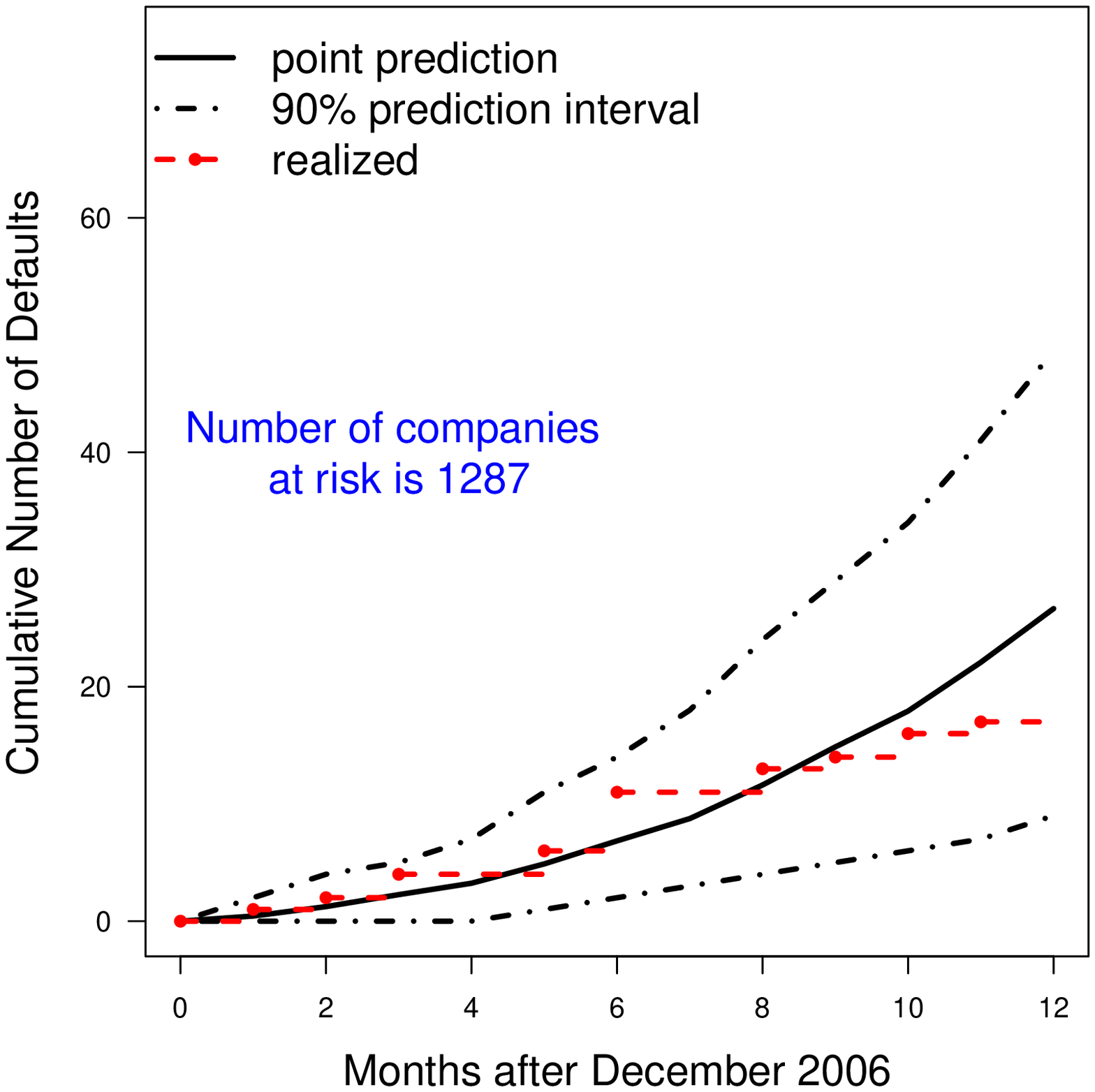}\\
(a) Defaults Predictions for 2006 & (b) Defaults Predictions for 2007\\
\includegraphics[width=0.45\textwidth]{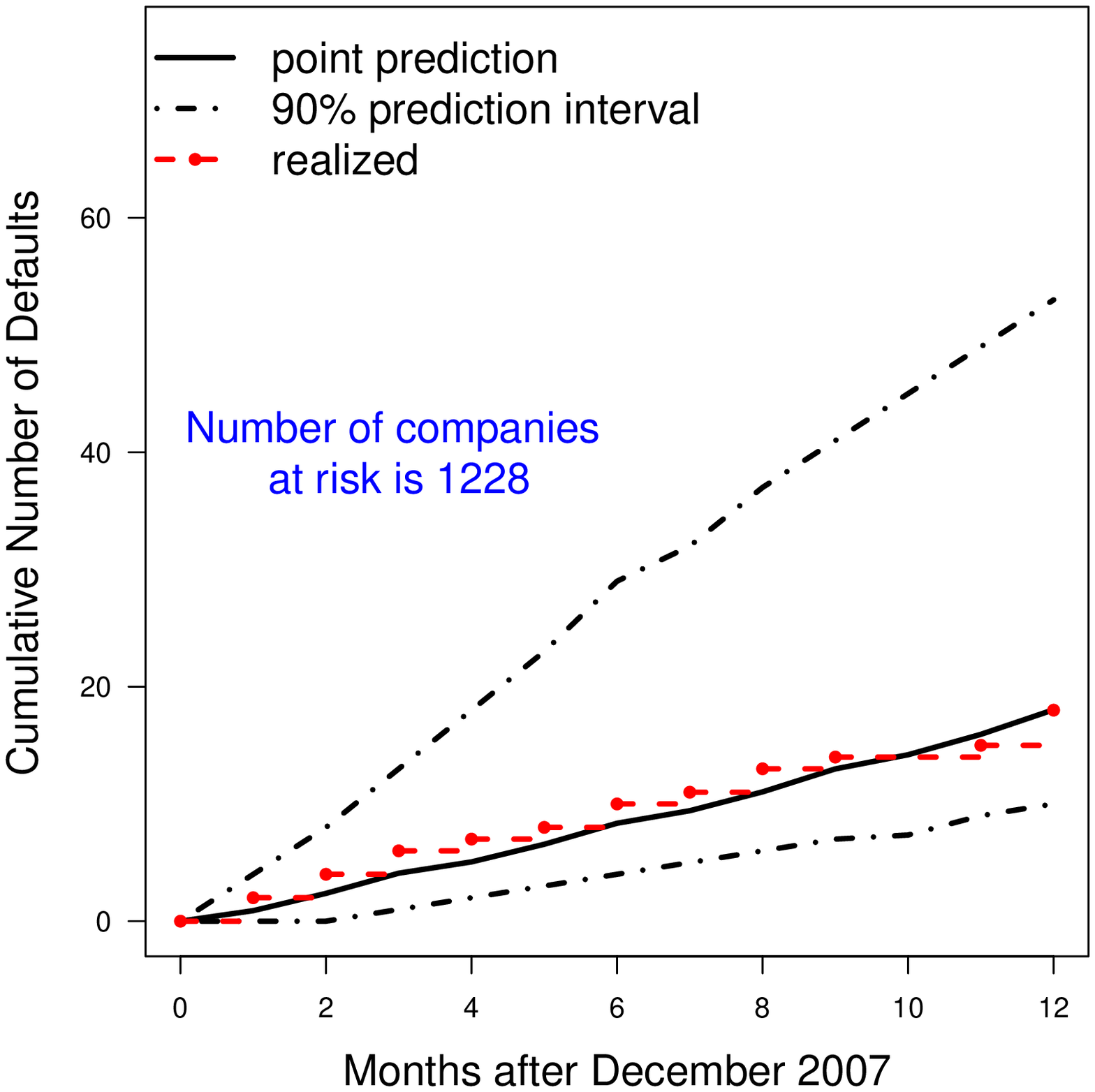}&
\includegraphics[width=0.45\textwidth]{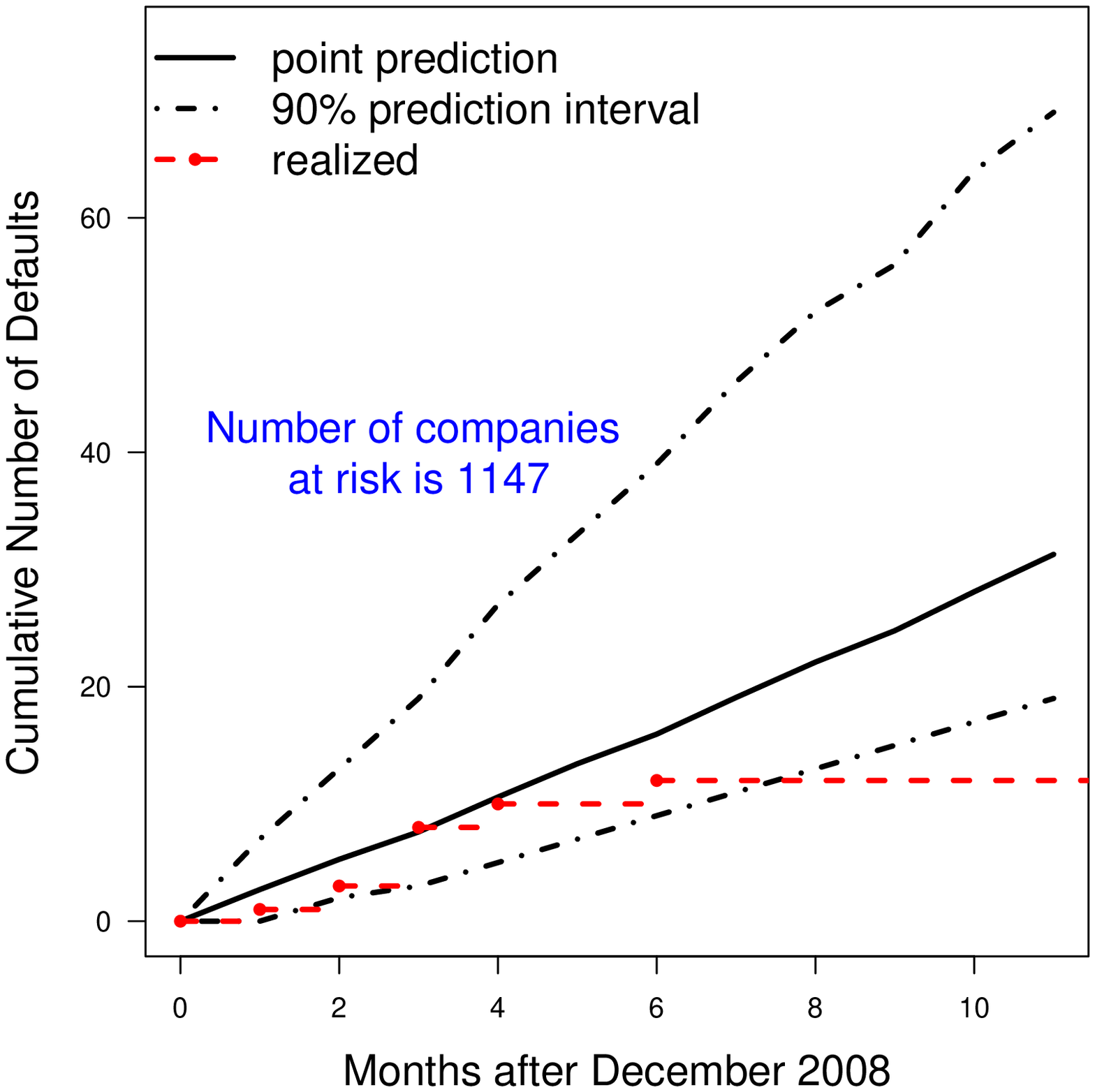}\\
(c) Defaults Predictions for 2008 & Defaults Predictions for 2009
\end{tabular}
\end{center}
\caption{Cumulative number of defaults in the one-year periods and the associated PI for all the units at risk.}\label{fig:popu.pi}
\end{figure}

We first consider  four respective one-year periods during 2006--2009, and for each period we conduct a one-year ahead prediction of the total number of defaults.
 We apply   \textbf{Algorithm 2} in Section \ref{agi} to obtain the prediction intervals for the total numbers.
%
%
%
Fig~\ref{fig:popu.pi} summarizes the results. Specifically for each one-year period,   Fig~\ref{fig:popu.pi} shows the predicted cumulative number of defaults  and the associated 90\% two-sided PI.  In each panel of Fig~\ref{fig:popu.pi}, the solid step plot with dots indicates the actual cumulative numbers of defaults by month, and the solid straight line represents the predicted average number of defaults.

From Fig~\ref{fig:popu.pi}, we find that the predictive assessments of the overall credit risk  levels are different between these four years by observing that, for example, the total predicted number of defaults in 2006 is much smaller than that of 2007.
Such an observation well matches the situations that actually happened.
Hence, for assessing the overall level of future credit risks, historical data are informative, and   it is an promising  evidence for using  quantitative methods to incorporate historical data information for future predictions.
We also note that Fig~\ref{fig:popu.pi} is reporting the cumulative number of predicted defaults, and the widening trend of the PIs reflect an  expected  fact that the level of uncertainties associated with predictions is increasing over time.  That is, one always needs to take more caution when applying predictions over a longer term due to higher level of associated uncertainties.

Additionally, we also see that predictions with the same model are not working equally well for all four one-year periods.
For 2007 and 2008, the mean predictions agrees well with the actual cumulative numbers of defaults, showing that the model works very well for these periods of time.
For 2006, the actual number of defaults is above the mean prediction but still falls between the mean prediction and the upper bounds of the PI (i.e., the 95\% percentile). 
Such an observation may indicate that the actual situation in 2006 was somewhat different from what happened before so that the model used was not able to perfectly reflect the future situation.  Nevertheless, the approach still performs reasonably well by observing the fact that the PI is narrow for 2006 and the actual number of defaults still fall with in the PI.
 In 2009, the mean predictions well matches  the actual cumulative number of defaults in the first six months.  However, as we can see from the figure, there is an abrupt change in the defaults occurrence by observing that no default was recorded in the second half of 2009, a phenomenon that may be due to the governmental interventions.
Our data used for prediction are only up to the beginning of 2009, thus predicting such an abruption is hard for a quantitative method.  Further investigation on  modifying the parametric modeling may be needed to incorporate sudden change of the market conditions.

\subsection{Individual Default Risk Predictions and Uncertainties}\label{sec:app.indi.pred}
We now present the performance of the point predictions and PIs for  future default probabilities of individual firms.  We  find from our studies that the level of uncertainties associated  with the point predictions, quantified by the width of PIs,  can  be highly informative in analyzing and predicting default risks.  

\begin{figure}
\begin{center}
\begin{tabular}{cc}
\includegraphics[height=0.2\textheight, width=0.4\textwidth]{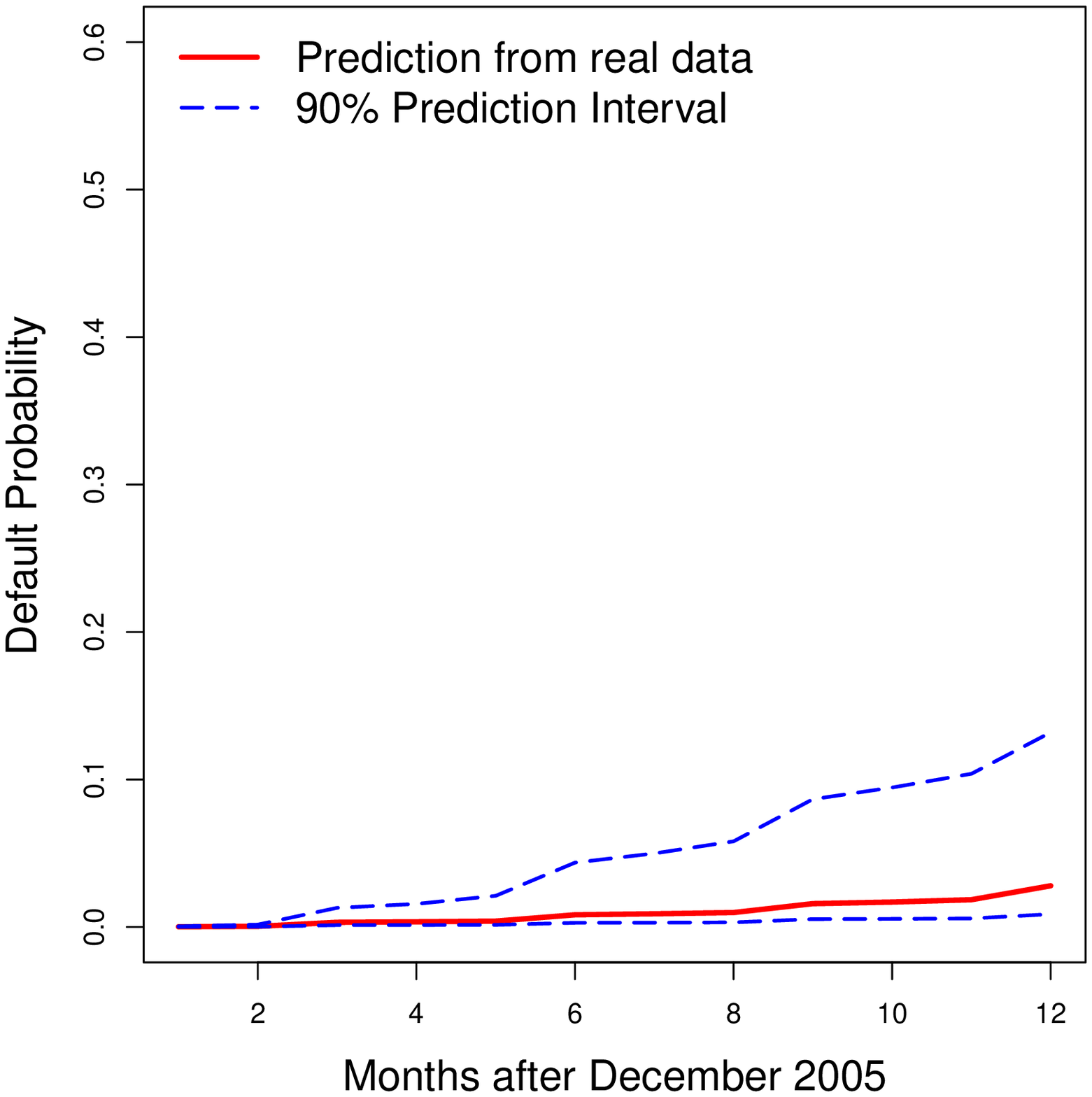}&
\includegraphics[height=0.2\textheight, width=0.4\textwidth]{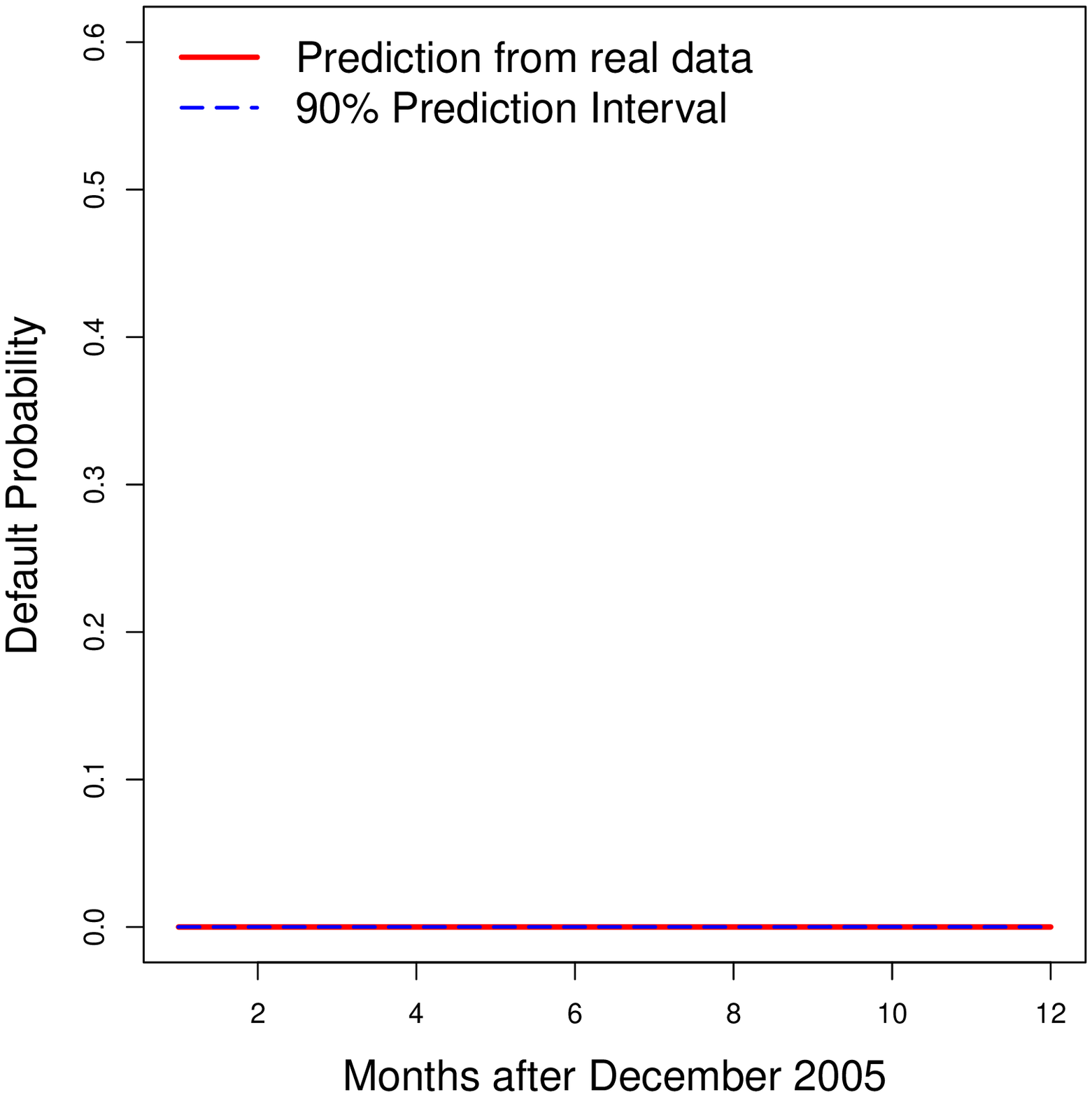}\\
(a) Emerge Interactive Inc & (b) Microsoft Corp \\
\includegraphics[height=0.2\textheight, width=0.4\textwidth]{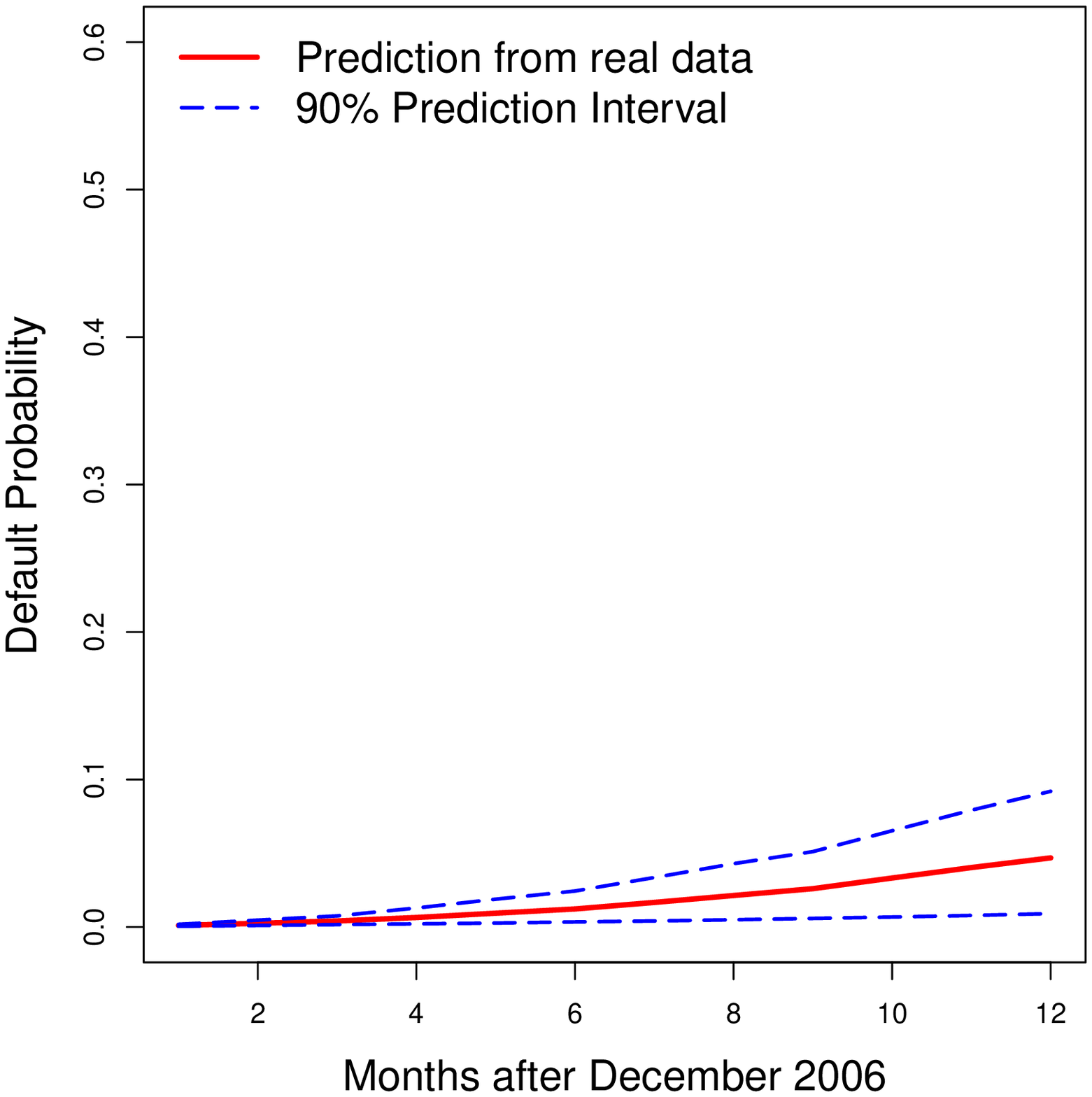}&
\includegraphics[height=0.2\textheight, width=0.4\textwidth]{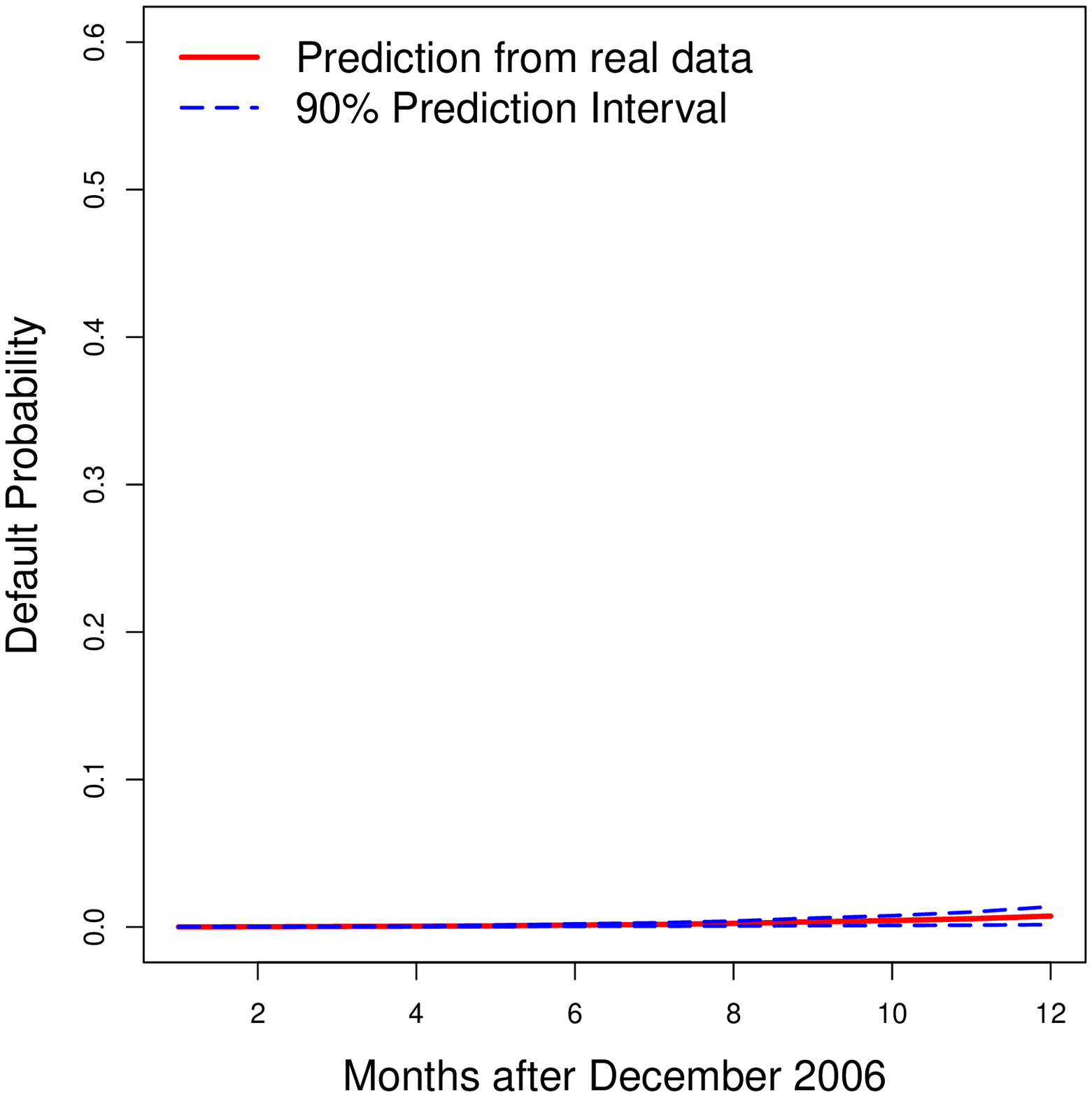}\\
(c) Lehman Brothers  & (d) City National Corp \\
\includegraphics[height=0.2\textheight, width=0.4\textwidth]{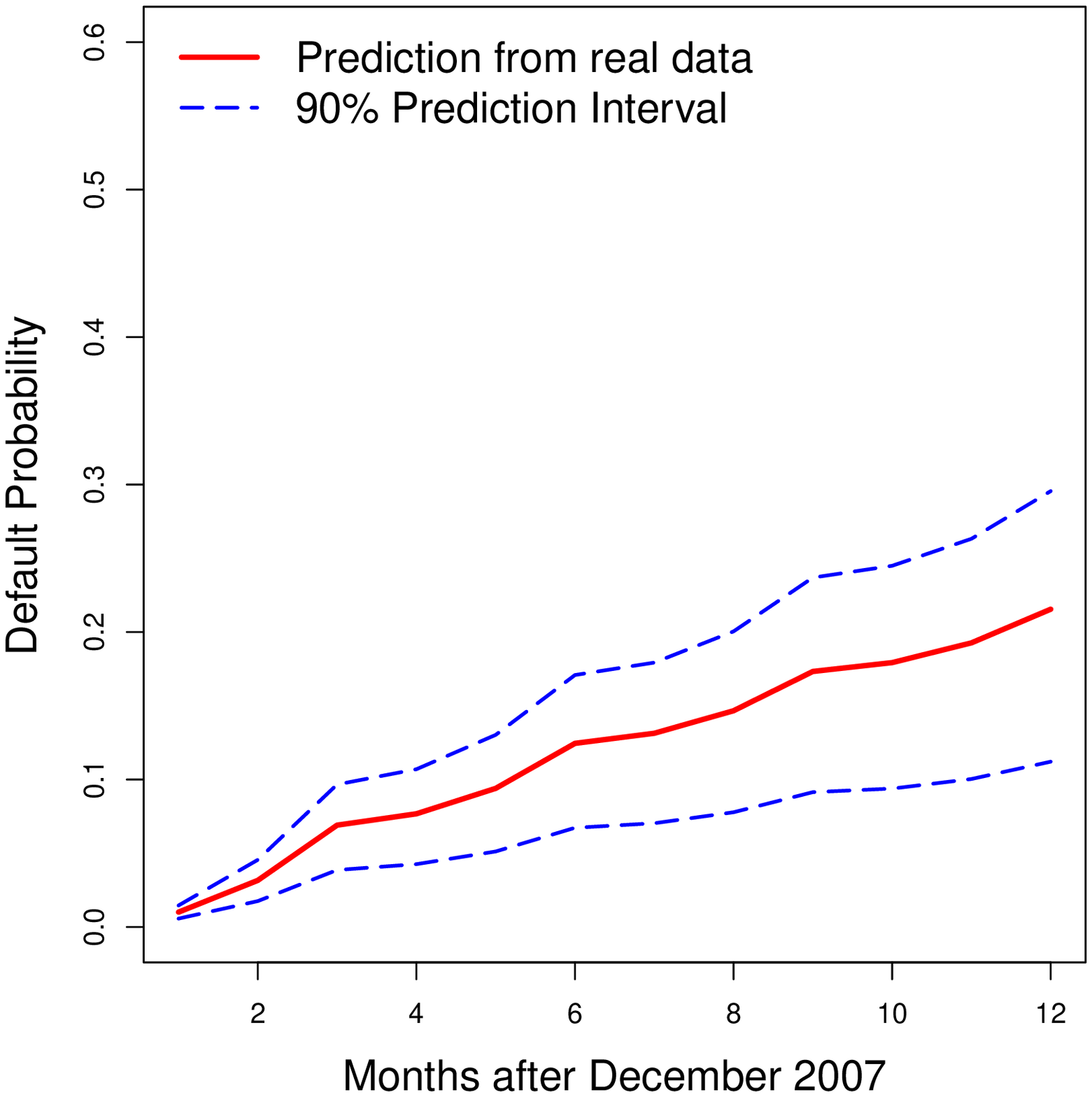}&
\includegraphics[height=0.2\textheight, width=0.4\textwidth]{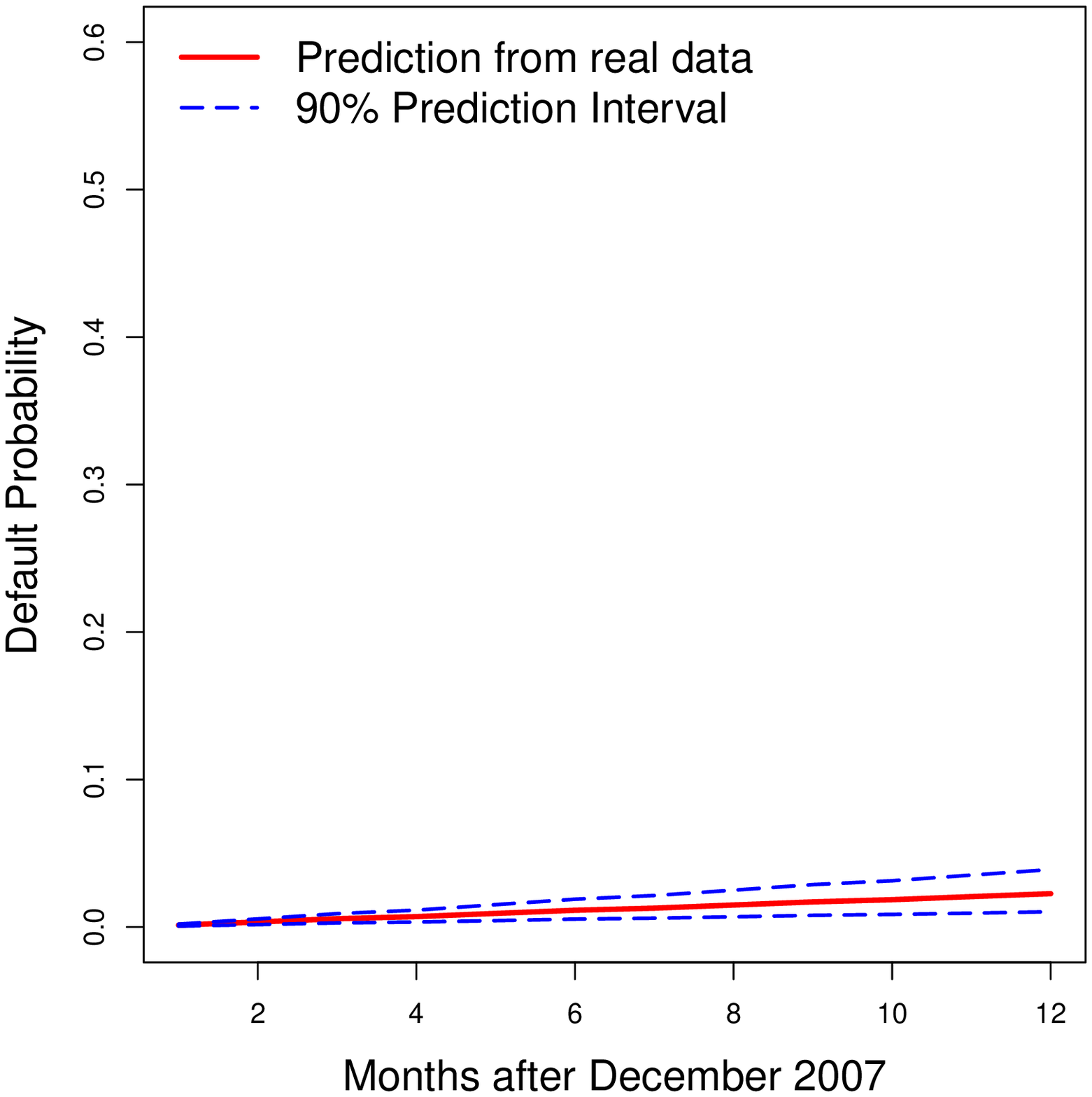}\\
(e) Bankunited Financial& (f) Bank of America Corp \\
\includegraphics[height=0.2\textheight, width=0.4\textwidth]{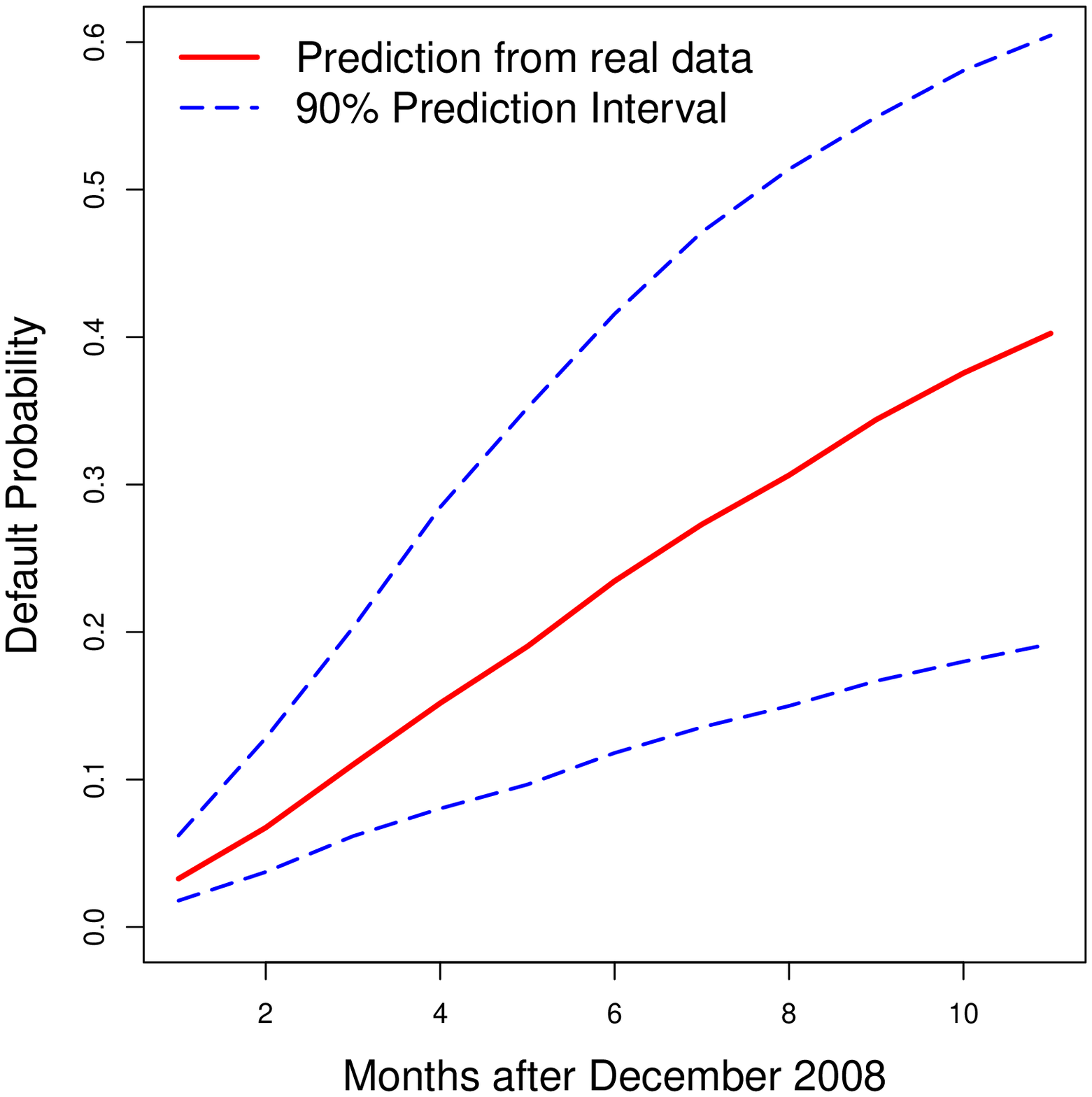}&
\includegraphics[height=0.2\textheight, width=0.4\textwidth]{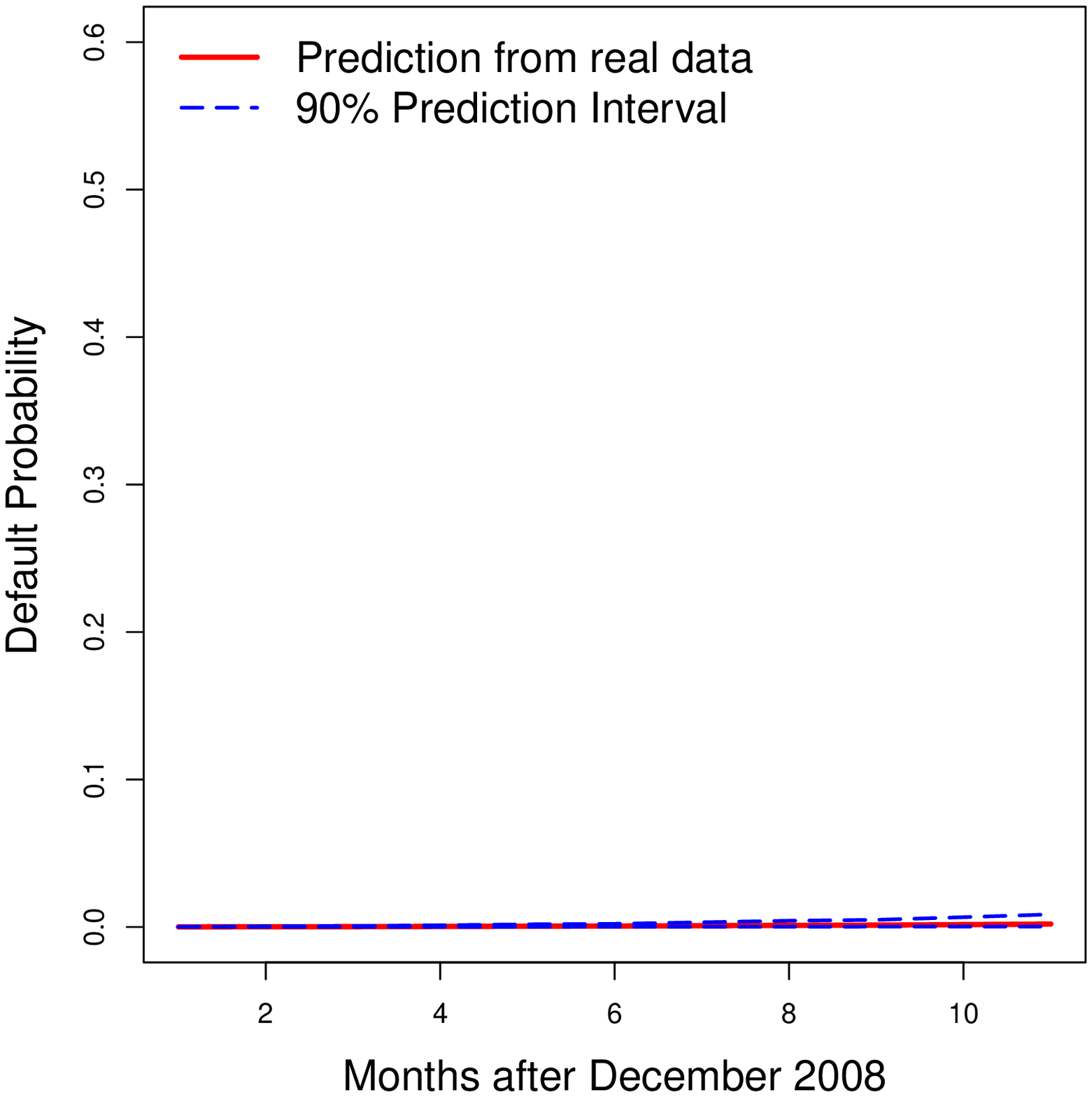}\\
(g) TierOne Corp & (h) Tellabs Inc
\end{tabular}
\end{center}
\caption{Predictions for individual default probabilities and the associated 90\% PI.}\label{fig:ind.fail}
\end{figure}

For the same four one-year periods (2006, 2007, 2008 and 2009) as in Section \ref{sec:app.popu.pred},
we assess each individual firm's default risks with monthly point predictions for its future default probabilities and the associated PIs using  \textbf{Algorithm 3} in Section \ref{sec:indi.pred}. The results for some selected firms are reported in Fig \ref{fig:ind.fail}.
We choose four companies that actually went default during the periods of time. For comparisons, we also present side by side another four companies from the same industrial sector but did not go default.  By presenting the results in the same scale, we clearly see striking differences between the profiles of the default predictions.  Specifically, those companies who actually went default are predicted to have substantially higher probabilities of going default. Additionally, the associated PIs are also much wider than those  companies who actually did not go default.

From the predictions for individual firms, it is promising to observe that the predicted default probabilities for those companies who actually went default are among the highest, providing us a crucial device for differentiating  companies based on quantitative credit risk assessments.
Emerge Interactive Inc., defaulted in 2006,  was a technology company providing food-safety, individual-animal tracking and supply-management services. According to the point default probability predictions at the end of 2006, Emerge Interactive Inc. has the highest default risk among all the five companies that actually defaulted in 2006.
Overall,  its predicted default probability is  ranked the 8th out of all 1,352 companies at risk. %
Lehman Brothers Holdings Inc., defaulted in 2007,  was the fourth largest investment bank in the US.  The predicted default probability   of Lehman Brothers in 2007 is ranked the 141st out of all 1287 companies at risks.
Bankunited Financial Corp,  defaulted in 2008,  was a savings and loan association.
Its predicted default probability is the 4th out of all 1228 companies at risk.
TierOne Corporation, defaulted in 2009,  was the holding company for TierOne Bank.  Its predicted default probability in 2009 is ranked the 9th out of all 1228 companies at risk.

Besides the level of default risks assessed by the point predictions, the associated PIs are providing information from a new dimension.  Visually, it is clear to see that the PIs are wider for those  actually defaulted companies.   Numerically,  for example,  the point prediction  for the probability of  Emerge Interactive Inc  going default in 2016 is 0.0347 and the associated 90\% PI is (0.0087, 0.1319).  In contrast, the counterparts for Microsoft are 0.00002 and (0, 0.00005), indicating striking differences between high risk and low risk companies.
As shown in the coming Section \ref{sec:Logit.reg}, we actually find that the level of uncertainties measured by the length of the PIs can provide extra information additional to the point predictions that can be potentially used for improving the accuracy of default predictions.

\subsection{Power Curves and Prediction Performances}\label{sec:ROC}
\begin{figure}
\begin{center}
\begin{tabular}{cc}
\includegraphics[width=0.45\textwidth]{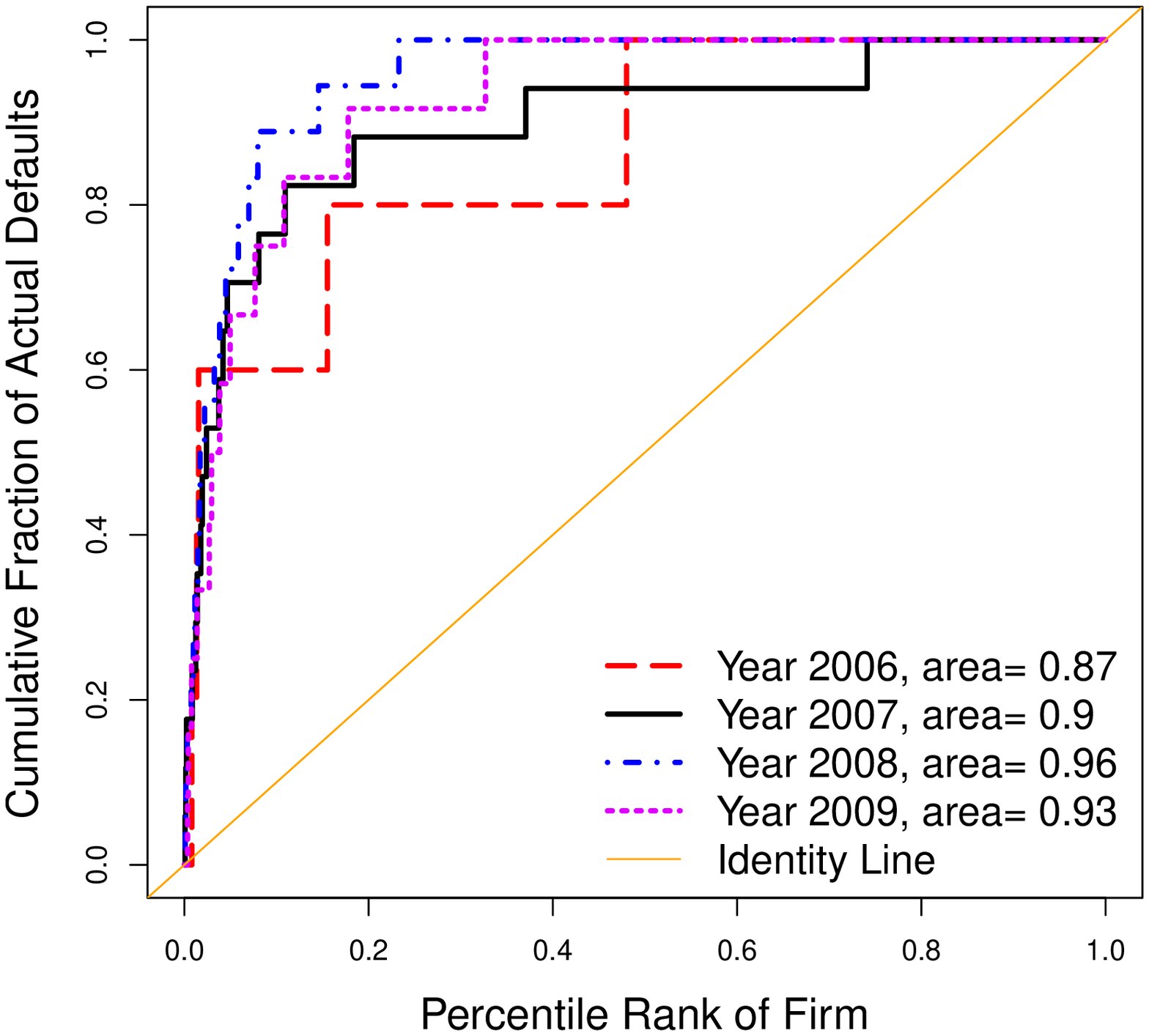}&
\includegraphics[width=0.45\textwidth]{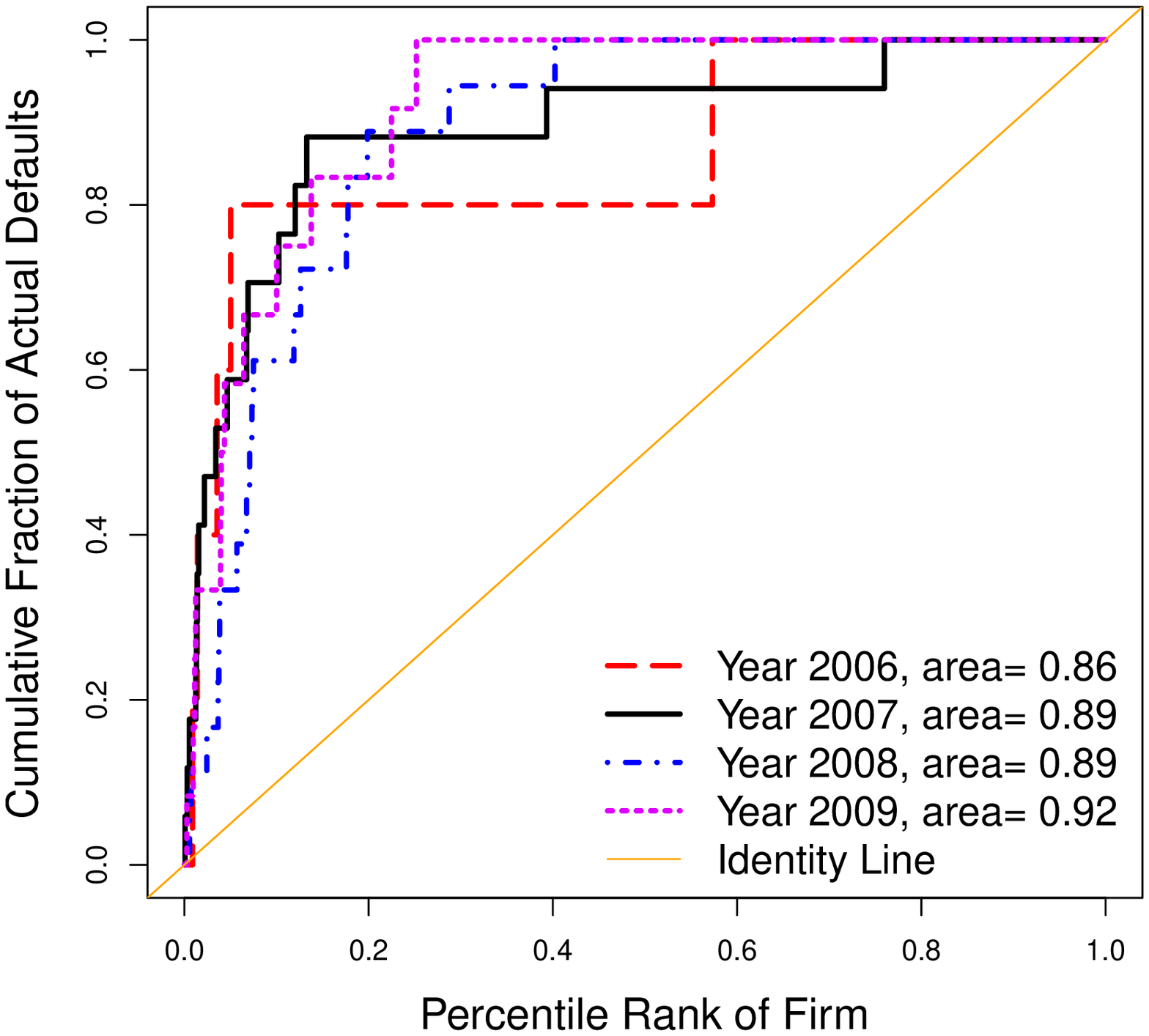}\\
(a) Point Prediction & (b) Width of PI
\end{tabular}
\end{center}
\caption{Out of sample prediction power curves by the point prediction and width of PI.}\label{fff}
\end{figure}

We now evaluate the out-of-sample default prediction performances for the four one-year periods.
For such a purpose, we plot the receiver operating characteristic (ROC) curves in Fig~\ref{fff},  which are also referred to as power curves in the literature, e.g., \cite{DuffieSaitaWang2007}.
A power curve is constructed  by plotting the cumulative fraction of actual defaults versus  the corresponding percentile the quantitative measure used to predictively rank all firms at risk.
That is, a steeply increasing curve is the evidence of good performance using the  corresponding ranking measure.   Equivalently,  larger area under the curve (AUC)  means better predictive performance.
Here, we consider two quantities  -- the predicted default probabilities and the lengths of the associated PIs -- for ranking all firms at risk to differentiate the defaulted firms.  Fig~\ref{fff}(a) and \ref{fff}(b) respectively show the power curves corresponding to these two quantities.

From Fig \ref{fff}, we can see that both predictive quantities have reasonable prediction performances, achieving AUCs near 0.9 out of  the maximum 1.
This again demonstrates the promising applications of quantitative methods for predicatively assess corporate default risks.
The predicted point default probability overall  performs slightly better than the width of the PI.
Since the width of PI is not intended for predicting future defaults, such an observation itself is interesting and informative and further research  on credit risks and their evaluations are needed for understanding such a phenomenon.
Moreover, we find that the width of PI is complementary to the point prediction of default probability; see Section \ref{sec:Logit.reg}.

\subsection{Default Predictions and Associated Uncertainties}\label{sec:Logit.reg}

We see from  Section \ref{sec:ROC} that point default probability predictions and the width of the associated PIs perform similarly effective for differentiating defaulted firms.
Then a natural question of interests is can the level of uncertainties measured by the width of PIs provide extra information for enhancing the prediction performance?
As an attempt to explore the answer for that question, we conduct a logistic regression with the default status as the response variable and both the predicted default probability and the width of the associated PIs as predictor.  Summary of the model fitting is reported in Table~\ref{tab:logit}.

As for the adequacy of the logistic regression model fitting, the deviance of the model is 578.61, while the deviance for the null model is  400.79. A chi-square test yields a $p$-value less than $0.001$, showing that the model is highly significant. For the overall goodness of fit,  we also did the Hosmer-Lemeshow test whose $p$-value is 0.5335, indicating that he the model provides a good fit to the data.  For comparison, we fit another logistic regression model dropping the length of the prediction interval from the model.  As a result, the Hosmer-Lemshow test \cite{Hosmer2013} of the reduced model has a $p$-value less than $0.001$, a significant evidence that the reduced model without the PI length is not adequate.

Table \ref{tab:logit} confirms that both larger  predicted default probability and wider PI   indicate  higher default risk. However, given the point prediction and the interaction between the two variables, the width of PI is no longer significantly associated with the default risk. 
Most interestingly,  however, a highly  significant interaction between the two predictors is detected  by the logistic regression, telling that using the PI width besides the point default predictions is  statistically informative.
Such an observation is quite reasonable from the perspective that one could be more confident in predicting a company's future default with smaller range of the prediction intervals.
That is, the width of PI has the potential of providing extra information for assessing  the  corporate default risks, suggesting an interesting topic for further investigations.

\begin{table}
\centering
\caption{Summary of the Logistic regression model.}
\begin{tabular}{lcccc}\hline\hline
 & Estimate & Std. Error & z value & Pr($>$$|$z$|$) \\
  \hline
 Intercept         & -6.1340 & 0.2898 & -21.1691 & $<0.0001$ \\
   PI width        &    2.0742 & 3.3577 & 0.6177 & 0.5367 \\
 Point prediction  &   49.6683 & 6.9704 & 7.1256 & $<0.0001$ \\
 PI width $\times$ Point prediction & -99.8712 & 14.4081 & -6.9316 & $<0.0001$ \\
   \hline\hline
\end{tabular}
\label{tab:logit}
\end{table}

\subsection{Model Diagnostics}

\begin{figure}
\begin{center}
\begin{tabular}{cc}
\includegraphics[width=0.45\textwidth, angle=270]{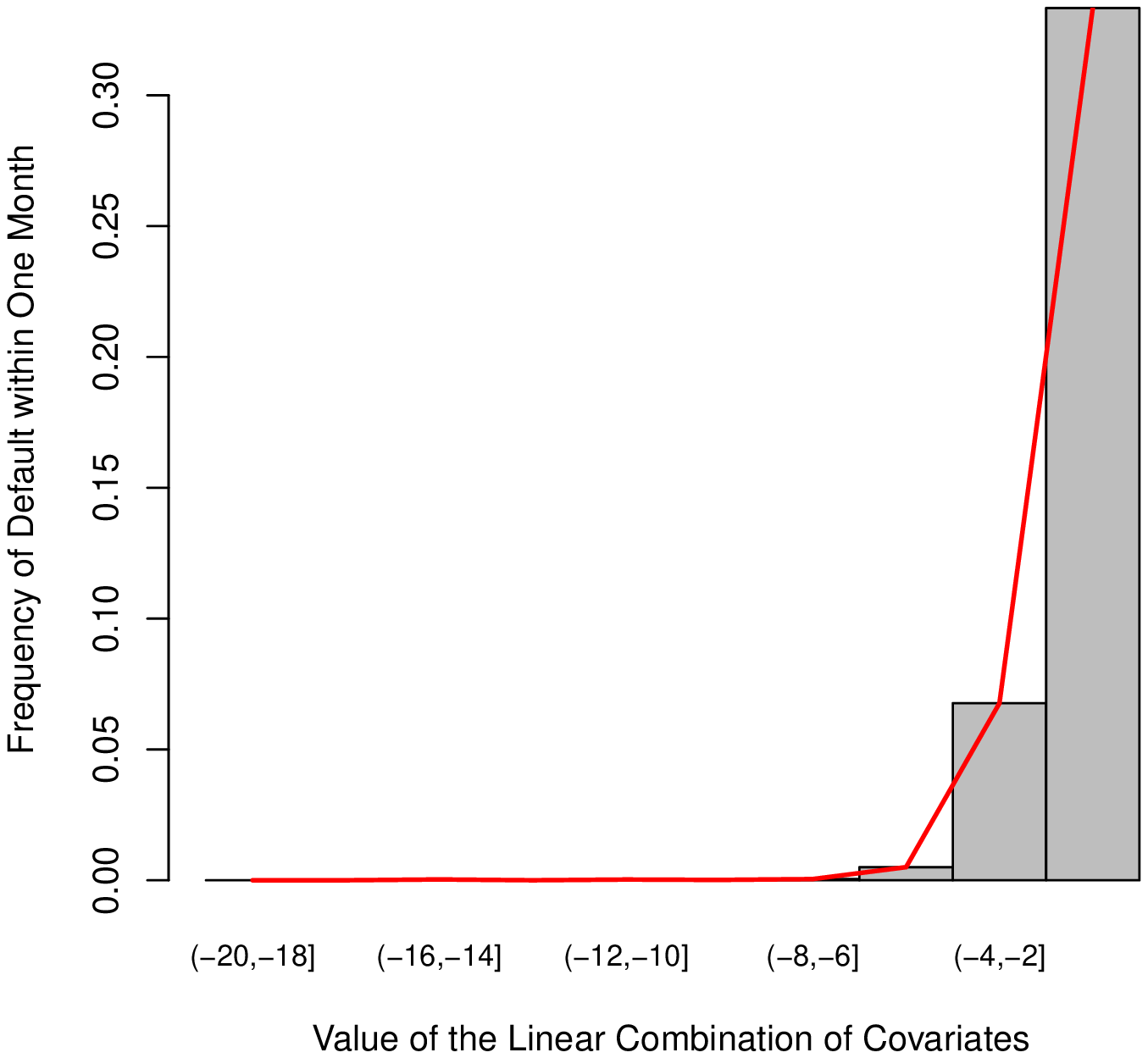}&
\includegraphics[width=0.45\textwidth, angle=270]{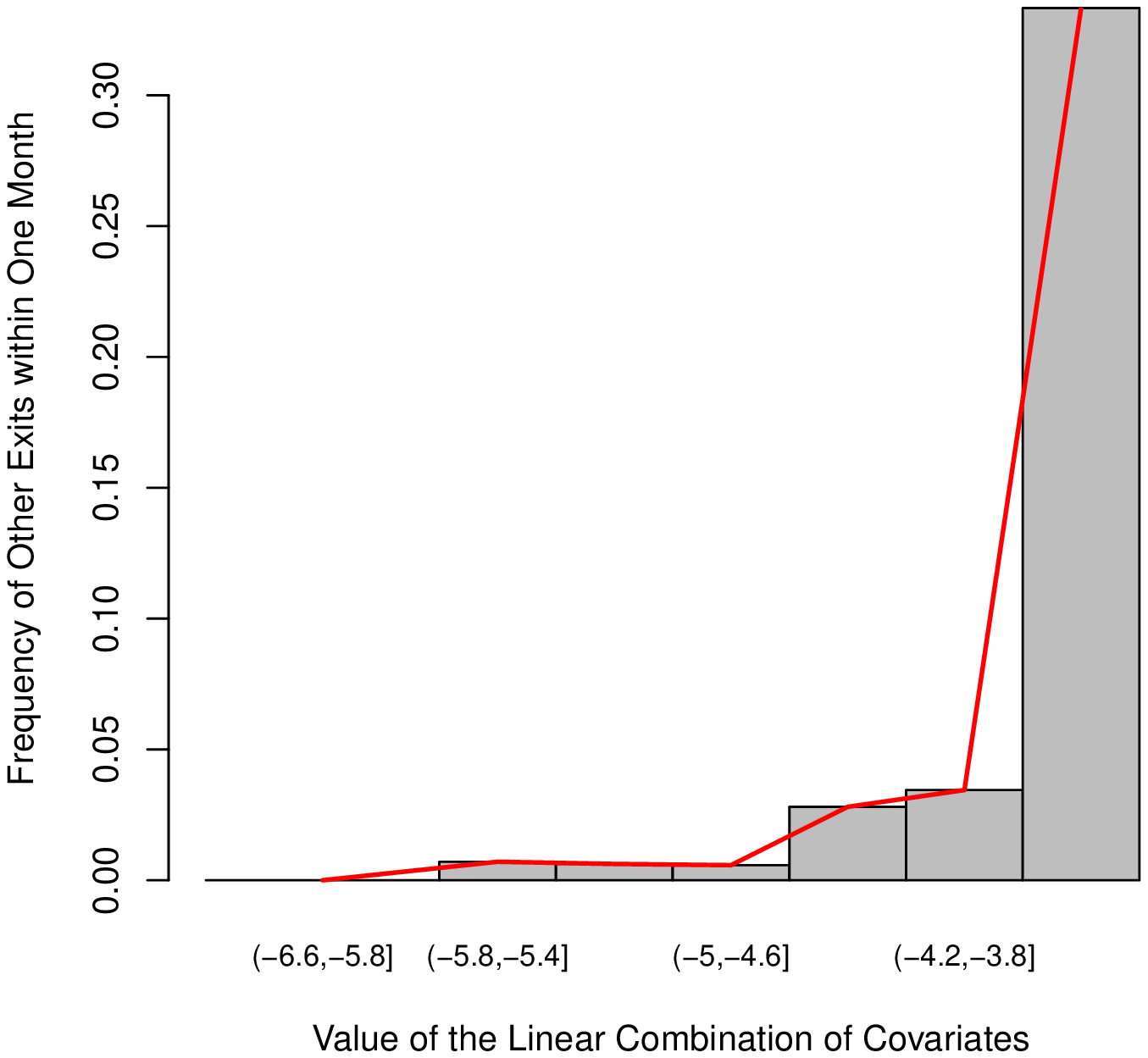}\\
(a) Defaults & (b) Other Exits
\end{tabular}
\end{center}
\caption{Empirical frequency of monthly defaults vs values of the estimated linear combinations broken down into ten intervals.  The red solid line is the estimated exponential function. }\label{f:d2}
\end{figure}

For assessing the adequacy of the model fitting to the data set,  we conduct some model diagnostics.
For assessing the effect of the dynamic factor model specified by (\ref{eq:facmodel}) and (\ref{eq:statem}), we attempted a fitting of a two-factor model but with no  dynamic structure (\ref{eq:statem}). As a result, we found that the dynamic model improves the fitting in the sense of reducing the mean residual sum of squares  by 10\%.  To check the exponential linear form of the intensity functions,  we calculate the estimated values of the specified  linear functions and break down the range of the values into ten intervals.  Then we aggregate the companies  according to ten intervals of the values of the estimated linear functions, and then obtain the respective total numbers  of  the companies.
The empirical frequencies of the defaults  respectively on each interval are reported by bars in Fig \ref{f:d2}(a), overlayed by the values of the exponential linear functions.
 The  shape of the  red line in Fig \ref{f:d2}(a) satisfactorily  validates exponential linear form of the intensity function for defaults.  
Similarly,  we obtain Fig \ref{f:d2}(b) and validates the exponential form of the intensity function for other types of exits.

\begin{figure}
\begin{center}
\begin{tabular}{cc}
\includegraphics[width=0.45\textwidth]{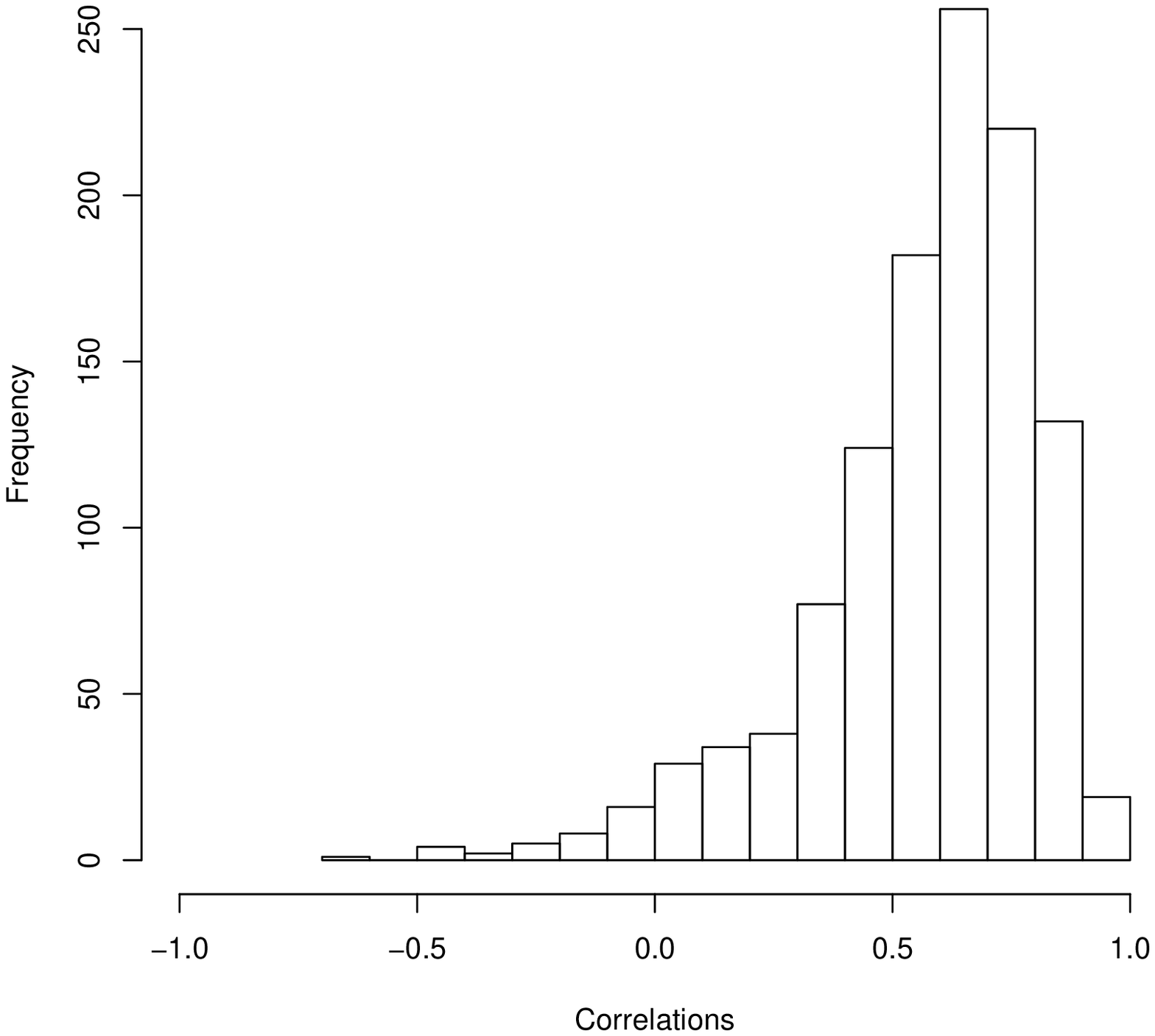}&
\includegraphics[width=0.45\textwidth]{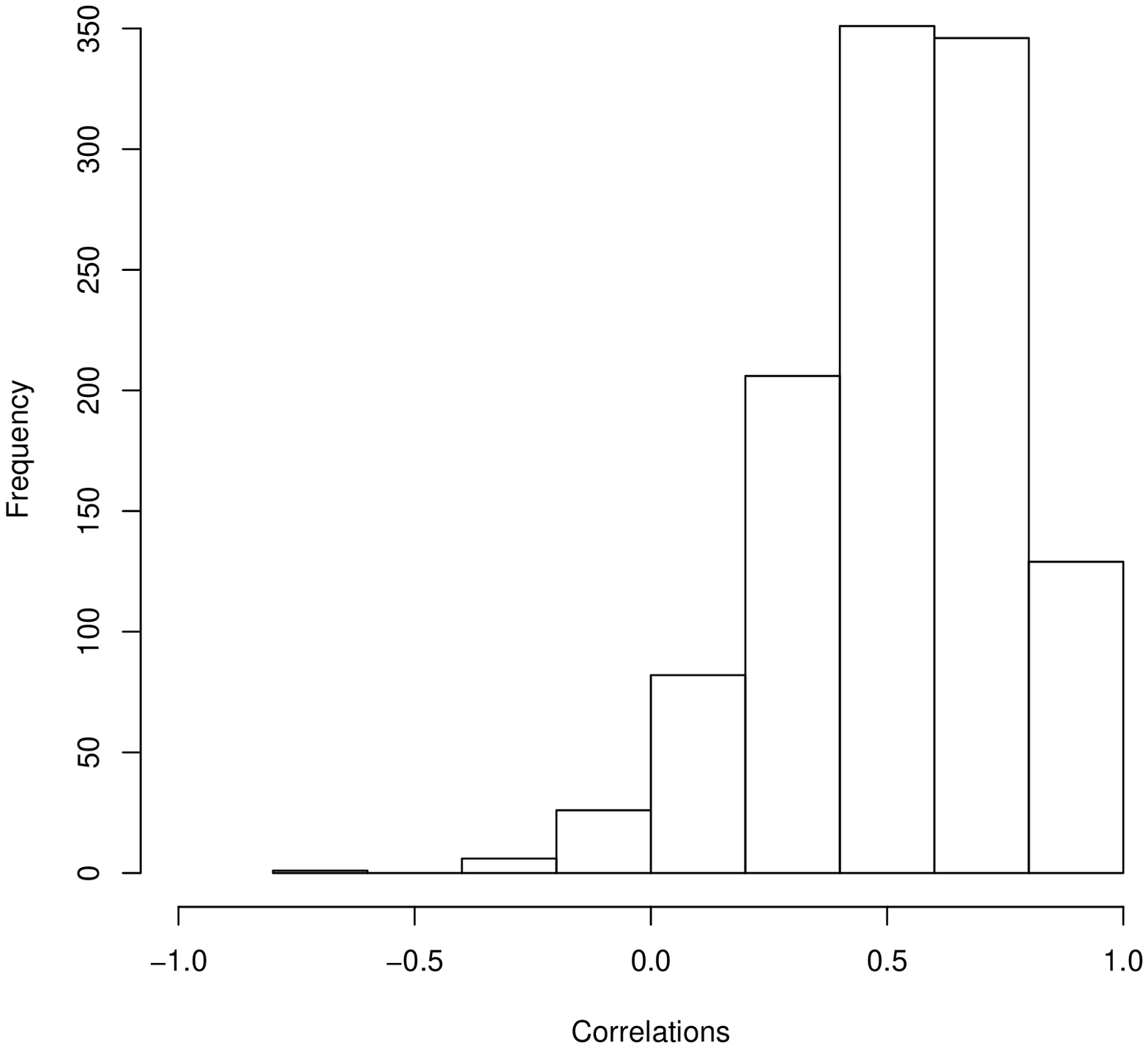}\\
(a) Distance to Default & (b) Trailing Return
\end{tabular}
\end{center}
\caption{Correlations between the fitted covariates and observed covariates for the two firm specific variables. }\label{f:d3}
\end{figure}

For checking the adequacy of the model for the covariate process,  we conducted some numerical and  graphical model diagnostics for checking  the goodness of fit. We compute the fitted covariate values according to the model and compare them with the actual observed values. In an ideal situation the fitted values and the observed values should display strong positive correlations.
Fig \ref{f:d3} shows the histograms of the correlations between the fitted covariates and the observed values for the two firm specific covariate.   From Fig \ref{f:d3}, we can see that a majority of the correlations are reasonably high, indicating an overall good fit of the models.
The average correlations for the distance to default and the trailing returns are 0.57 and $0.52$ respectively.
 Given the large number of individual companies, some lack of fit  is inevitable,  and some dedicated further adjustment for the models can also be possible.   The fitting of the covariate model for  the two macro economic variables is also reasonably good.
The correlations between the fitted values and the observed values are 0.56 and 0.82 respectively for the returns of the Treasury bill and the SP 500 index.
Overall, the model for the covariates modeling is flexible and fit the data reasonably well considering that we use one model for covariates of so many companies.

\section{A Simulation Study} \label{sim}

The simulation setting is constructed based on the data set of the Section 4 on  US corporate defaults from 1990 to 2009.  Specifically, we  first take a random sample of size $n$ from the set of the companies that are at risk as of the end of year 2008.  We vary $n\in \{400,600,800,1000\}$  to assess the impact from the scale of the problem on the accuracy of the framework.   Intuitively with the same period of time,  the larger the $n$ is, the more difficult it is to predict future defaults.

Upon selecting the $n$ companies, we generate  both the events of default and other types of exist based on the parametric intensity models outlined in Sections 2.3 and 4.2.  To ensure reasonable numbers of events in the simulations and for simplicity, we set the same intensity functions for both events of defaults and other exists, particularly, with the same parameters $\beta=(-5.26,0.1,-1.2, -0.045, -0.084)^{\T}$.  These values closely mimic those estimated from the real data set as reported in Table 1.

We then estimate the parameters of the covariate model outlined in Section 2.4 based on the $n$ companies, and fixed the values of the parameters throughout the simulations with size  $n$.  With the estimated parameters, we simulate the  monthly differenced values of the covariate from the same model as outlined in Section 2.4.  Then we generate the covariate process   from the simulated differenced values,  with the first observations  of the companies taken from that  of the random sample of size $n$.
Then, with the simulated covariate process, we generate both events of defaults and other exits with the parametric intensity functions specified  as described earlier.

\begin{figure}
\begin{center}
\begin{tabular}{cc}
\includegraphics[width=0.45\textwidth]{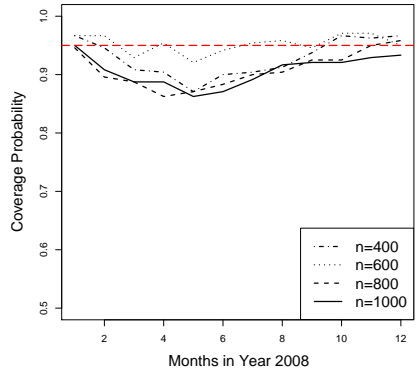}
&\includegraphics[width=0.45\textwidth]{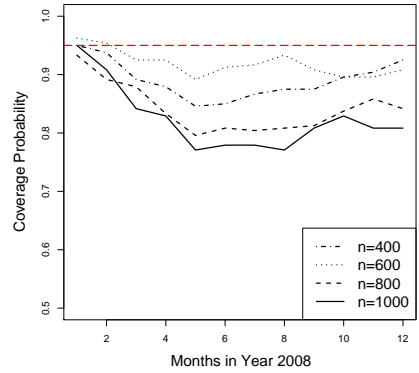} \\
(a) 95\% calibrated PI & (b) 95\%  uncalibrated PI  \\
\includegraphics[width=0.45\textwidth]{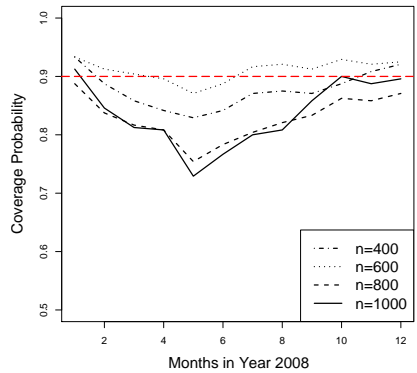}&
\includegraphics[width=0.45\textwidth]{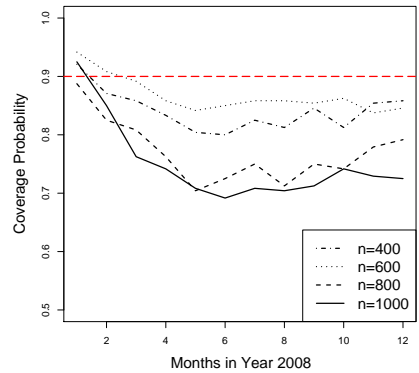}\\
(c) 90\% calibrated PI & (d) 90\%  uncalibrated PI
\end{tabular}
\end{center}
\caption{Coverage probability of calibrated and uncalibrated PI from simulation study.}\label{fig:cp.cali.ncali}
\end{figure}

Upon generating the simulated data set with both the covariate process and the time-to-events, we apply our method to produce prediction intervals of  the cumulative number of defaults in each month of 2008.  We also compare the two types of prediction intervals -- calibrated and uncalibrated ones -- as outlined in Section 3.2.
The simulations for each  sample size $n$ are repeated for 240 times. The results of the accuracy in terms of the percentage that the prediction intervals cover the true cumulative default numbers are reported in Fig \ref{fig:cp.cali.ncali}.

From Fig \ref{fig:cp.cali.ncali}, we have a few observations.
First and foremost on the empirical accuracy of the uncertainties assessment, 
we observe that the empirical frequencies of the prediction coverages are close to the nominal levels for all multiple period predictions when $n$ is smaller.  When $n$ is larger, the coverage of the PI is also very accurate  for the cumulative default counts  within a closer time horizon from the origin of predictions.  Second, we see that  the coverages generally get worse for larger time horizon cumulative predictions.  Since we are examining the cumulative predictions, the  main reason should be the error aggregations in the cumulative counts predictions. That is, even the  coverages of the prediction intervals for the number of events at  each  individual month are close to the nominal level,  the performance of the PI for cumulative counts  will still be more off the target due to  that all errors  are aggregating together.
Given the same amount of information, this also reflects the practical difficulty in obtaining accurate predictions for longer time horizons.
Third, the calibrated intervals perform substantially better than the uncalibrated ones, indicating the merits of applying calibrated procedures for prediction intervals.  
Since the naive prediction intervals (without calibration) only capture the intrinsic randomness in the random variables and ignores the uncertainty from parameter estimations,  they tend to be narrow so that the coverage tends to be smaller than the nominal level.  In contrast, the advantage of calibration relies on the fact that  it takes the additional source of  uncertainties into account.  Hence the coverage of the calibrated prediction intervals tend to be closer to the nominal level. 

%

\section{Discussions and Future Work}\label{sec:con2}

We consider the challenging  problem of assessing uncertainties associated with corporate default predictions by carefully disentangling and quantifying the contributing sources for the point predictions.
%
An application of our framework to a large-scale US Corporate data set shows that our point predictions have good out-of-sample performance, and is promising in quantifying the uncertainties in predictions.  Our framework also helps for a better understanding of the default mechanism by providing an additional dimension of insights from assessing the level of uncertainties associated with point predictions.
%
%
%
%
%
%
With limited access to a powerful  modern computational  facility (160 hours in parallel on 80 CPUs), we accomplish the tasks for assessing the uncertainties associated with corporate default predictions,  demonstrate the feasibility for  solving  this long overdue  important  large scale practical problem  with many challenging practical features.
Our study reveals high level of uncertainties associated default predictions especially when conducting longer term predictions. We believe that is  mainly due to the nature of the prediction problem involving unknown future dynamics of those factors affecting the default mechanism.  Hence more cautions are necessary when using quantitative tools for mid- and long-term default predictions.   We also note that the rationale of the quantitative modeling is to predict future events based on historical information.
Thus for all model based quantitative predictions, they should be interpreted as that if the market conditions are consistent with the historical scenarios, then the predictions are valid. Otherwise,  one has to take serious cautions.

Further investigations for assessing the uncertainties associated with corporate default risk predictions are clearly desirable.
Our framework needs parametric models for the default mechanism and the covariate processes. Extending the scope of the framework and  evaluating its robustness are clearly important.  For example, other methods dealing with the default mechanism may be considered.
Additional considerations on features such as the cycling effect \cite{Koopman2005} and systemic risk \cite{GieseckeKim_2011_MS} can also be investigated.
 Interesting questions  also include how to efficiently incorporate more variables, and how  a violation of the parametric models may affect the accuracy of the assessed level of uncertainties.
There is another important consideration on the correlations between defaults.
Recent investigations reveal the correlations and even clustering effect of the occurrences of corporate defaults; see, among others,  the frailty modeling approaches with random effects in \cite{Duffieetal2009},   the dynamic frailty modeling approaches of \cite{Koopman2008a} and \cite{Koopman2011}, and the jump cumulative default intensity function approach of \cite{Peng2009}.   It will also be interesting to explore the impact from the random and even clustering effects in constructing the prediction intervals.
We hope to conduct further investigations on these problems in future.
Both theoretical and practical investigations are also needed for exploring the high-dimensional covariate process for large number of companies with the global market data.
Instead of resorting to the simulation based approach we take, a potential direction for assessing uncertainties associated with default predictions could be applying the Bayesian approaches that being capable of producing the posterior distribution of the quantities of interests. How to develop parsimonious and effective Bayesian modeling, and how to design and implement efficient practical computational  framework are interesting and open challenging problems.

\section*{Acknowledgment}
 We thank  Professor  Brendan Murphy,   the Associate Editor and three referees for their insightful comments and constructive suggestions that have greatly improved the paper.  The authors acknowledge Advanced Research Computing at Virginia Tech for providing computational resources. We are very thankful to Dr Tao Wang for his generous help on the data sets.   Tang gratefully acknowledges the support from a Grant from the Risk Management Institute of  National University of Singapore. 

\begin{supplement}
\sname{Supplement Material}\label{suppA}
\stitle{Supplement to ``Disentangling  and Assessing Uncertainties in Multiperiod  Corporate Default Risk Predictions''}
\slink[url]{http://www.e-publications.org/ims/support/dowload/imsart-ims.zip}
\sdescription{Detail of the EM algorithm. }
\end{supplement}


\begin{thebibliography}{}

\bibitem[\protect\citeauthoryear{Altman, Resti, and Sironi}{Altman
  et~al.}{2004}]{Altman2004}
Altman, E., A.~Resti, and A.~Sironi (2004).
\newblock Default recovery rates in credit risk modelling: A review of the
  literature and empirical evidence.
\newblock {\em Economic Notes\/}~{\em 33\/}(2), 183--208.

\bibitem[\protect\citeauthoryear{Altman}{Altman}{1968}]{Altman_1968_JF}
Altman, E.~I. (1968).
\newblock Financial ratios, discriminant analysis, and the prediction of
  corporate bankruptcy.
\newblock {\em Journal of {F}inance\/}~{\em 23}, 589--609.

\bibitem[\protect\citeauthoryear{Bai and Ng}{Bai and Ng}{2008}]{BaiNg2008}
Bai, J. and S.~Ng (2008).
\newblock Large dimensional factor analysis.
\newblock {\em Foundations and Trends in Econometrics\/}~{\em 3}, 89--163.

\bibitem[\protect\citeauthoryear{Bai and Wang}{Bai and
  Wang}{2015}]{BaiWang2012}
Bai, J. and P.~Wang (2015).
\newblock Identification and bayesian estimation of dynamic factor models.
\newblock {\em Journal of Business and Economic Statistics\/}~{\em 33},
  221--240.

\bibitem[\protect\citeauthoryear{Banbura and Modugno}{Banbura and
  Modugno}{2012}]{BanburaModugno2012}
Banbura, M. and M.~Modugno (2012).
\newblock Maximum likelihood estimation of factor models on datasets with
  arbitrary pattern of missing data.
\newblock {\em Journel of Applied Econometrics\/}~{\em 29}, 133--160.

\bibitem[\protect\citeauthoryear{Beaver}{Beaver}{1966}]{Beaver1966}
Beaver, W.~H. (1966).
\newblock Financial ratios as predictors of failure.
\newblock {\em Journal of Accounting Research\/}~{\em 4}, 71.

\bibitem[\protect\citeauthoryear{Beaver}{Beaver}{1968}]{Beaver1968}
Beaver, W.~H. (1968).
\newblock Market prices, financial ratios, and the prediction of failure.
\newblock {\em Journal of Accounting Research\/}~{\em 6\/}(2), 179.

\bibitem[\protect\citeauthoryear{Beaver, Correia, and McNichols}{Beaver
  et~al.}{2012}]{Beaver2012}
Beaver, W.~H., M.~Correia, and M.~F. McNichols (2012).
\newblock Do differences in financial reporting attributes impair the
  predictive ability of financial ratios for bankruptcy?
\newblock {\em Review of Accounting Studies\/}~{\em 17\/}(4), 969--1010.

\bibitem[\protect\citeauthoryear{Bharath and Shumway}{Bharath and
  Shumway}{2008}]{Bharath2008}
Bharath, S.~T. and T.~Shumway (2008).
\newblock Forecasting default with the merton distance to default model.
\newblock {\em Review of Financial Studies\/}~{\em 21\/}(3), 1339--1369.

\bibitem[\protect\citeauthoryear{B{\"o}hning, Dietz, Schaub, Schlattmann, and
  Lindsay}{B{\"o}hning et~al.}{1994}]{Bohningetal1994}
B{\"o}hning, D., E.~Dietz, R.~Schaub, P.~Schlattmann, and B.~G. Lindsay (1994).
\newblock The distribution of the likelihood ratio for mixtures of densities
  from the one-parameter exponential family.
\newblock {\em Annals of the Institute of Statistical Mathematics\/}~{\em 46},
  373--388.

\bibitem[\protect\citeauthoryear{Br{\"a}uning and Koopman}{Br{\"a}uning and
  Koopman}{2014}]{Braeuning2014}
Br{\"a}uning, F. and S.~J. Koopman (2014).
\newblock Forecasting macroeconomic variables using collapsed dynamic factor
  analysis.
\newblock {\em International Journal of Forecasting\/}~{\em 30\/}(3), 572--584.

\bibitem[\protect\citeauthoryear{Campbell, Hilscher, and Szilagyi}{Campbell
  et~al.}{2008}]{Campelletal_2008_JF}
Campbell, J.~Y., J.~Hilscher, and J.~Szilagyi (2008).
\newblock In search of distress risk.
\newblock {\em Journal of Finance\/}~{\em 63}, 2899--2939.

\bibitem[\protect\citeauthoryear{Chava and Jarrow}{Chava and
  Jarrow}{2004}]{Chava2004}
Chava, S. and R.~A. Jarrow (2004).
\newblock Bankruptcy prediction with industry effects.
\newblock {\em Review of Finance\/}~{\em 8\/}(4), 537--569.

\bibitem[\protect\citeauthoryear{Diebold and Mariano}{Diebold and
  Mariano}{1995}]{DieboldMariano_1995_JBES}
Diebold, F. and R.~S. Mariano (1995).
\newblock Comparing predictive accuracy.
\newblock {\em Journal of Business and Economics Statistics\/}~{\em 13},
  253--263.

\bibitem[\protect\citeauthoryear{Ding, Tian, Yu, and Guo}{Ding
  et~al.}{2012}]{Dingetal_2012_JASA}
Ding, A., S.~Tian, Y.~Yu, and H.~Guo (2012).
\newblock A class of discrete transformation survival models with application
  to default probability prediction.
\newblock {\em Journal of the American Statistical Association\/}~{\em 107},
  990--1003.

\bibitem[\protect\citeauthoryear{Duan, Sun, and Wang}{Duan
  et~al.}{2012}]{DuanSunWang2012}
Duan, J., J.~Sun, and T.~Wang (2012).
\newblock Multiperiod corporate default prediction -- a forward inensity
  approach.
\newblock {\em Journal of Econometrics\/}~{\em 170}, 191--209.

\bibitem[\protect\citeauthoryear{Duffie}{Duffie}{2011}]{Duffie2011}
Duffie, D. (2011).
\newblock {\em Measuring Corporate Default Risk (Clarendon Lectures in
  Finance)}.
\newblock Oxford University Press.

\bibitem[\protect\citeauthoryear{Duffie, Eckner, Horel, and Saita}{Duffie
  et~al.}{2009}]{Duffieetal2009}
Duffie, D., A.~Eckner, G.~Horel, and L.~Saita (2009).
\newblock Frailty correlated default.
\newblock {\em The Journal of Finance\/}~{\em 64}, 2089--2123.

\bibitem[\protect\citeauthoryear{Duffie and Lando}{Duffie and
  Lando}{2001}]{Duffie2001}
Duffie, D. and D.~Lando (2001, may).
\newblock Term structures of credit spreads with incomplete accounting
  information.
\newblock {\em Econometrica\/}~{\em 69\/}(3), 633--664.

\bibitem[\protect\citeauthoryear{Duffie, Saita, and Wang}{Duffie
  et~al.}{2007}]{DuffieSaitaWang2007}
Duffie, D., L.~Saita, and K.~Wang (2007).
\newblock Multi-period corporate default prediction with stochastic covariates.
\newblock {\em Journal of Financial Economics\/}~{\em 83}, 635--665.

\bibitem[\protect\citeauthoryear{Durbin and Koopman}{Durbin and
  Koopman}{2012}]{Durbin2012}
Durbin, J. and S.~J. Koopman (2012).
\newblock {\em Time Series Analysis by State Space Methods: Second Edition
  (Oxford Statistical Science Series)}.
\newblock Oxford University Press.

\bibitem[\protect\citeauthoryear{Giesecke and Kim}{Giesecke and
  Kim}{2011}]{GieseckeKim_2011_MS}
Giesecke, K. and B.~Kim (2011).
\newblock Systemic risk: what defaults are telling us.
\newblock {\em Management Science\/}~{\em 57}, 1387--1405.

\bibitem[\protect\citeauthoryear{Giesecke, Longstaff, Schafer, and
  Strebulaev}{Giesecke et~al.}{2011}]{Gieseckeetal_2011_JFE}
Giesecke, K., F.~Longstaff, S.~Schafer, and I.~Strebulaev (2011).
\newblock Corporate bond default risk: a 150-year perspective.
\newblock {\em Journal of Financial Economics\/}~{\em 102}, 233--250.

\bibitem[\protect\citeauthoryear{Hillegeist, Keating, Cram, and
  Lundstedt}{Hillegeist et~al.}{2004}]{Hillegeist2004}
Hillegeist, S.~A., E.~K. Keating, D.~P. Cram, and K.~G. Lundstedt (2004, mar).
\newblock Assessing the probability of bankruptcy.
\newblock {\em Review of Accounting Studies\/}~{\em 9\/}(1), 5--34.

\bibitem[\protect\citeauthoryear{Hong}{Hong}{2013}]{Hong2013}
Hong, Y. (2013).
\newblock On computing the distribution function for the poisson binomial
  distribution.
\newblock {\em Computational Statistics and Data Analysis\/}~{\em 59}, 41--51.

\bibitem[\protect\citeauthoryear{Hong, Meeker, and McCalley}{Hong
  et~al.}{2009}]{HongMeekerMcCalley2009}
Hong, Y., W.~Q. Meeker, and J.~D. McCalley (2009).
\newblock Prediction of remaining life of power transformers based on left
  truncated and right censored lifetime data.
\newblock {\em The Annals of Applied Statistics\/}, 857--879.

\bibitem[\protect\citeauthoryear{Hosmer, Lemeshow, and Sturdivant}{Hosmer
  et~al.}{2013}]{Hosmer2013}
Hosmer, D.~W., S.~Lemeshow, and R.~X. Sturdivant (2013).
\newblock {\em Applied Logistic Regression}.
\newblock John Wiley and Sons Ltd.

\bibitem[\protect\citeauthoryear{Kalbfleisch and Prentice}{Kalbfleisch and
  Prentice}{2002}]{KalbfleischPrentice2002}
Kalbfleisch, J.~D. and R.~L. Prentice (2002).
\newblock {\em The Statistical Analysis of Failure Time Data\/} (2nd ed.).
\newblock John Wiley and Sons Inc.

\bibitem[\protect\citeauthoryear{Koopman and Lucas}{Koopman and
  Lucas}{2005}]{Koopman2005}
Koopman, S.~J. and A.~Lucas (2005).
\newblock Business and default cycles for credit risk.
\newblock {\em Journal of Applied Econometrics\/}~{\em 20\/}(2), 311--323.

\bibitem[\protect\citeauthoryear{Koopman and Lucas}{Koopman and
  Lucas}{2008}]{Koopman2008}
Koopman, S.~J. and A.~Lucas (2008).
\newblock A non-gaussian panel time series model for estimating and decomposing
  default risk.
\newblock {\em Journal of Business \& Economic Statistics\/}~{\em 26\/}(4),
  510--525.

\bibitem[\protect\citeauthoryear{Koopman, Lucas, and Monteiro}{Koopman
  et~al.}{2008}]{Koopman2008a}
Koopman, S.~J., A.~Lucas, and A.~Monteiro (2008).
\newblock The multi-state latent factor intensity model for credit rating
  transitions.
\newblock {\em Journal of Econometrics\/}~{\em 142\/}(1), 399--424.

\bibitem[\protect\citeauthoryear{Koopman, Lucas, and Schwaab}{Koopman
  et~al.}{2011}]{Koopman2011}
Koopman, S.~J., A.~Lucas, and B.~Schwaab (2011).
\newblock Modeling frailty-correlated defaults using many macroeconomic
  covariates.
\newblock {\em Journal of Econometrics\/}~{\em 162\/}(2), 312--325.

\bibitem[\protect\citeauthoryear{Koopman, Lucas, and Schwaab}{Koopman
  et~al.}{2012}]{Koopman2012}
Koopman, S.~J., A.~Lucas, and B.~Schwaab (2012).
\newblock Dynamic factor models with macro, frailty, and industry effects for
  us default counts: the credit crisis of 2008.
\newblock {\em Journal of Business \& Economic Statistics\/}~{\em 30\/}(4),
  521--532.

\bibitem[\protect\citeauthoryear{Lam and Yao}{Lam and
  Yao}{2012}]{LamYao_2012_AOS}
Lam, C. and Q.~Yao (2012).
\newblock Factor modelling for high-dimensional time series: inference for the
  number of factors.
\newblock {\em The Annals of Statistics\/}~{\em 40}, 694--726.

\bibitem[\protect\citeauthoryear{Lawless and Fredette}{Lawless and
  Fredette}{2005}]{LawlessFredette2005}
Lawless, J.~F. and M.~Fredette (2005).
\newblock Frequentist prediction intervals and predictive distributions.
\newblock {\em Biometrika\/}~{\em 92}, 529--542.

\bibitem[\protect\citeauthoryear{Meeker and Escobar}{Meeker and
  Escobar}{1998}]{meekerescobar1998}
Meeker, W.~Q. and L.~A. Escobar (1998).
\newblock {\em Statistical Methods for Reliability Data}.
\newblock New York: John Wiley \& Sons, Inc.

\bibitem[\protect\citeauthoryear{Merton}{Merton}{1974}]{Merton1974}
Merton, R.~C. (1974).
\newblock On the pricing of corporate debt: the risk structure of interest
  rates.
\newblock {\em Journal of Finance\/}~{\em 29}, 449--470.

\bibitem[\protect\citeauthoryear{Ohlson}{Ohlson}{1980}]{Ohlson1980}
Ohlson, J.~A. (1980).
\newblock Financial ratios and the probabilistic prediction of bankruptcy.
\newblock {\em Journal of Accounting Research\/}~{\em 18\/}(1), 109.

\bibitem[\protect\citeauthoryear{Pan and Yao}{Pan and
  Yao}{2008}]{PanYao_2008_Bioka}
Pan, J. and Q.~Yao (2008).
\newblock Modelling multiple time series via common factors.
\newblock {\em Biometrika\/}~{\em 95}, 365--379.

\bibitem[\protect\citeauthoryear{Peng and Kou}{Peng and Kou}{2009}]{Peng2009}
Peng, X. and S.~Kou (2009).
\newblock Default clustering and valuation of collateralized debt obligations.
\newblock {\em Working Paper\/}.

\bibitem[\protect\citeauthoryear{Schwaab, Koopman, and Lucas}{Schwaab
  et~al.}{2017}]{Schwaab2017}
Schwaab, B., S.~J. Koopman, and A.~Lucas (2017).
\newblock Global credit risk: World, country and industry factors.
\newblock {\em Journal of Applied Econometrics\/}~{\em 32\/}(2), 296--317.

\bibitem[\protect\citeauthoryear{Shumway}{Shumway}{2001}]{Shumway2001}
Shumway, T. (2001).
\newblock Forecasting bankruptcy more accurately: A simple hazard model.
\newblock {\em The Journal of Business\/}~{\em 74\/}(1), 101--124.

\bibitem[\protect\citeauthoryear{Stock and Watson}{Stock and
  Watson}{2002}]{stock2002}
Stock, J.~H. and M.~W. Watson (2002).
\newblock Forecasting using principal components from a large number of
  predictors.
\newblock {\em Journal of the American statistical association\/}~{\em
  97\/}(460), 1167--1179.

\bibitem[\protect\citeauthoryear{Tsay}{Tsay}{2010}]{Tsay_2010}
Tsay, R. (2010).
\newblock {\em Analysis of Financial Time Series\/} (3rd ed.).
\newblock New York: Wiley.

\bibitem[\protect\citeauthoryear{Tsay}{Tsay}{2013}]{Tsay2013}
Tsay, R.~S. (2013).
\newblock {\em Multivariate Time Series Analysis: With R and Financial
  Applications}.
\newblock Wiley.


\bibitem[\protect\citeauthoryear{Yuan, Tang, Hong, and Yang}{Yuan et~al.}{2018}]{Yuanetal2018}
Yuan, M.,  Tang, C.Y., Hong, Y., and Yang, J. (2018).
\newblock Supplement to ``Disentangling  and Assessing Uncertainties in Multiperiod  Corporate Default Risk Predictions''. 


\bibitem[\protect\citeauthoryear{Volkova}{Volkova}{1996}]{Volkova1996}
Volkova, A.~Y. (1996).
\newblock A refinement of the central limit theorem for sums of independent
  random indicators.
\newblock {\em Theory of Probability and Its Applications\/}~{\em 40},
  791--794.

\bibitem[\protect\citeauthoryear{Zmijewski}{Zmijewski}{1984}]{Zmijewski1984}
Zmijewski, M.~E. (1984).
\newblock Methodological issues related to the estimation of financial distress
  prediction models.
\newblock {\em Journal of Accounting Research\/}~{\em 22}, 59.

\end{thebibliography}
\end{document}